\documentclass[useAMS,usenatbib]{mn2e}

\usepackage{amsmath}
\usepackage{amssymb}
\usepackage[dvipsnames]{xcolor}
\usepackage{graphicx}
\usepackage{graphics}
\usepackage{hyperref}
\hypersetup{
     colorlinks   = true,
     citecolor    = blue
}

\def\apj{{ ApJ}}
\def\apjl{{ApJL}}
\def\apjs{{ ApJS}}

\def\aap{{ A\&A}}

\def\mnras{{ MNRAS}}
\def\araa{{ ARA\&A}}

\def\nat {{ Nature}}
\def\pasj{{ PASJ}}

\def\physrep{{ Physics Reports}}

\def\prd{{ Phys. Rev. D}}

\def\prl{{ Phys. Rev. Lett}}

\def\nar{{New Astronomy Reviews}}
\def\mr{\mathrm}

\def\d{\mr{d}}

\def\mc{\mathcal}
\def\msun{M_{\rm \odot}}

\def\rg{r_{\rm g}}
\def\rd{r_{\rm d}}

\def\Md{M_{\rm d}}
\def\Mej{M_{\rm ej}}

\def\dotMd{\dot{M}_{\rm d}}
\def\dotMBH{\dot{M}_{\rm BH}}
\def\tvis{t_{\rm vis}}
\def\OmgK{\Omega_{\rm K}}

\def\Ye{Y_{\rm e}}
\def\me{m_{\rm e}}
\def\mp{m_{\rm p}}
\def\cs{c_{\rm s}}
\def\kB{k_{\rm B}}
\def\Phid{\Phi_{\rm d}}
\def\PhiBH{\Phi_{\rm BH}}
\def\Bp{B_{\rm p}}
\def\tevap{t_{\rm evap}}

\def\ue{u_{\rm e}}
\def\be{\beta_{\rm e}}
\def\ge{\gamma_{\rm e}}
\def\Ee{E_{\rm e}}

\def\swift{{\textit{Swift }}}

\newcommand{\lrb}[1]{\left({#1}\right)}
\newcommand{\lrsb}[1]{\left[{#1}\right]}
\newcommand{\lara}[1]{\left\langle{#1}\right\rangle}
\newcommand{\abs}[1]{\left|#1\right|}
\newcommand{\circled}[1]{\raisebox{.5pt}{\textcircled{\raisebox{-.9pt}{#1}}}}

\newcommand{\myemail}{wenbinlu@berkeley.edu}

\title[NS merger disk evolution]{Late-time accretion in neutron star mergers: implications for short gamma-ray bursts and kilonovae}
\author[Lu \& Quataert]
  {Wenbin Lu$^{1, 2}$\thanks{\myemail} and Eliot Quataert$^{2}$\\
  $^1$Departments of Astronomy and Theoretical Astrophysics Center, UC Berkeley, Berkeley, CA 94720, USA\\
  $^2$Department of Astrophysical Sciences, Princeton University, Princeton, NJ 08544, USA}

\begin{document}
\label{firstpage}
\maketitle

\begin{abstract}
We study the long-term ($t\gg 10\rm\, s$) evolution of the accretion disk after a neutron star(NS)-NS or NS-black hole merger, taking into account the radioactive heating by r-process nuclei formed in the first few seconds. We find that the cumulative heating eventually exceeds the disk's binding energy at $t\sim 10^2\mr{\,s}\, (\alpha/0.1)^{-1.8}(M/2.6\,\msun)^{1.8}$ after the merger, where $\alpha$ is the Shakura-Sunyaev viscosity parameter and $M$ is the mass of the remnant object. This causes the disk to evaporate rapidly and the jet power to shut off. We propose that this is the cause of the steep flux decline at the end of the extended emission (EE) or X-ray plateau seen in many short $\gamma$-ray bursts (GRBs). The shallow flux evolution before the steep decline is consistent with a plausible scenario where the jet power scales linearly with the disk mass. We suggest that the jets from NS mergers have two components --- a short-duration narrow one corresponding to the prompt $\gamma$-ray emission and a long-lasting wide component producing the EE. This leads to a prediction that ``orphan EE'' (without the prompt $\gamma$-rays) may be a promising electromagnetic counterpart for NS mergers observable by future wide-field X-ray surveys.
The long-lived disk produces a slow ejecta component that can efficiently thermalize the energy carried by $\beta$-decay electrons up to $t\sim 100\rm\, d$ and contributes $\sim\!10\%$ of the kilonova's bolometric luminosity at these late epochs. We predict that future ground-based and JWST near-IR spectroscopy of nearby ($\lesssim 100\rm\, Mpc$) NS mergers will detect narrow ($\Delta v\sim 0.01c$) line features a few weeks after the merger, which provides a powerful probe of the atomic species formed in these events.

\end{abstract}

\begin{keywords}
gamma-ray bursts: general --- hydrodynamics --- neutron star mergers --- black hole - neutron star mergers --- nuclear reactions, nucleosynthesis, abundances --- gravitational waves --- accretion, accretion discs
\end{keywords}

\section{Introduction}
The overall picture of GW170817/AT2017gfo, as inferred from its gravitational wave (GW) emission, $\gamma$-ray flash, kilonova, and broadband afterglow from a relativistic off-axis jet, confirmed the conjecture that binary neutron star (NS) mergers are a significant contributor to the r-process nucleosynthesis as well as the sources of short gamma-ray bursts \citep[GRBs,][]{1989Natur.340..126E, 1992ApJ...395L..83N, 2017ApJ...848L..13A, 2017ApJ...848L..14G, 2017Sci...358.1556C, 2017Sci...358.1559K, mooley18_VLBI_proper_motion, metzger19_kilonova_review, nakar20_GW170817_review, margutti21_GW170817_ARAA}.

During the first few seconds after the merger (roughly corresponding to the prompt $\gamma$-ray emission phase in a short GRB), the accretion disk is efficiently cooled by neutrino emission at a high accretion rate of $0.01$ to $0.1\rm\,\msun\,s^{-1}$ \cite[e.g.][]{1999ApJ...518..356P,2001ApJ...557..949N, 2002ApJ...579..706D, kohri05_NDAF}, and the gas near the disk mid-plane maintains a low electron fraction of $Y_{\rm e}\sim 0.1$ since electrons are mildly degenerate \citep{2007ApJ...657..383C, siegel18_GRMHD_disk}.

As a result of viscous evolution, the outer disk expands with time to larger radii. When the majority of the disk mass reaches a distance of a few hundred gravitational radii of the central object, where the temperature drops below a few MeV, some dramatic changes occur \citep{2007NJPh....9...17L, metzger08_onezone_model, 2008AIPC.1054...51B, metzger09_Ye_freeze}: (i) the disk is no-longer neutrino cooled and the heat generated by viscous dissipation and nuclear reactions is advected with the fluid motion; (ii) electrons become non-degenerate causing the electron fraction $Y_{\rm e}$ to increase; (iii) photo-disintegration becomes less efficient and free nucleons recombine to form He and heavier elements; (iv) the energy release of a few to $10\,$MeV/nucleon from nuclear recombination is sufficient to unbind a large fraction of the disk mass. Thus, we expect that the disk experiences a state transition with violent outflows, as has recently been shown in the general relativistic magnetohydrodynamic (GRMHD) simulations by \citet{siegel18_GRMHD_disk, fernandez19_long_GRMHD}.

The dynamical evolution of the electron fraction $Y_{\rm e}$ is controlled  by disk expansion on the viscous timescale $t_{\rm vis}$ and the timescale $t_{\rm weak}$ for pair capture on free nucleons (and hence the $n\leftrightarrow p$ conversion). If $t_{\rm vis}>t_{\rm weak}$, then the local fluid is close to $\beta$-equilibrium as the disk slowly expands and the electron fraction gradually increases to the equilibrium value of $Y_{\rm e}\simeq 0.5$ (no r-process nuclei will be generated in this case). However, simulations generally found $t_{\rm vis}\ll t_{\rm weak}$ at the time when He recombination occurs in the disk and subsequent nucleosynthesis starts, so the disk and outflow composition after the $\Ye$ freeze-out is on average at least mildly neutron-rich with $Y_{\rm e}\lesssim 0.4$ \citep[e.g.,][]{metzger09_Ye_freeze, fernandez13_disk_evolution_Ye}. Observationally, the 10 day-long, ``red'' kilonova emission in GW170817 requires a relatively massive (a few percent of $\msun$), neutron-rich ($\Ye\lesssim 0.25$) ejecta component, which is most likely ejected from the accretion disk \citep{shibata17_GW170817_Mej, metzger19_kilonova_review}.

Global numerical simulations of the post-merger accretion disk evolution \citep[e.g.,][]{siegel18_GRMHD_disk, 2018ApJ...869L..35R, fernandez19_long_GRMHD, murguia-berthier21_GRMHD_NDAF} only covered a timescale $t\lesssim 10\rm\, s$ and hence have not been able to determine the final fate of the disk at $t\gg 10\rm\,s$. The goal of this paper is to explore this late stage of evolution.

Since the material in the remaining disk after nuclear recombination is expected to be enriched with r-process nuclei (as a result of low $\Ye\lesssim 0.4$), heating due to radioactive decay may be dynamically important, for the following reason. The heating rate can be approximated as a power-law $q_{\rm h}\propto t^{-1.3}$ in the time interval of $10 \lesssim t \lesssim 10^4\rm\,s$ \citep[][see also Appendix \ref{sec:heating_rate}]{metzger10_kilonova, roberts11_kilonova_tidal_tail, 2015ApJ...815...82L, barnes16_heating_thermalization, 2017MNRAS.468...91H, 2018A&A...615A.132R}. The cumulative energy deposition from radioactive heating $\propto t q_{\rm h}\propto t^{-0.3}$ will eventually exceed the specific binding energy $\propto GM/r_{\rm  d}\propto t^{-2/3}$ (since the disk radius $\rd$ increases as $t^{2/3}$ from Kepler's third law). When this occurs, the disk material becomes unbound and is expected to quickly evaporate.

On the observational side, a significant fraction ($\gtrsim 1/4$) of short GRBs show in their $\gamma$/X-ray lightcurves an extended emission (EE) component or a plateau/shallow-decay phase lasting for about $10^2\rm\, s$ followed by a rapid decline steeper than $t^{-3}$ \citep{norris06_sGRB_EE, 2006ApJ...647.1213O, perley09_sGRB_EE, sakamoto11_BAT_catalog, gompertz13_sGRB_EE, 2013MNRAS.430.1061R, kaneko15_sGRB_EE}. Some low-redshift apparently longer-duration GRBs from which a supernova can be ruled out (and hence a NS merger scenario is favored) are dominated by an EE-like episode lasting for about $10^2\rm\, s$, e.g., GRB 060614, GRB 211211A, GRB 211227A \citep{gehrels06_GRB060614, galyam06_GRB060614, lu22_GRB211227A, rastinejad22_GRB211211A, yang22_GRB211211A}. The EE and X-ray plateau require a long-lived central engine over a timescale that is about two orders of magnitude longer than the duration of the prompt $\gamma$-ray emission in short GRBs \citep{berger14_sGRB_review}. We note that such a shallow-decay phase followed by a sharp drop is also found in the X-ray lightcurves some long GRBs, although the duration is of the order $10^4\rm\, s$ \citep[e.g.,][]{2007ApJ...670..565L, troja07_longGRB_EE, lyons10_longGRB_plateau}. The origin(s) of the EE and the X-ray plateau followed by a sudden flux decline are still debated.

Long GRBs are generally believed to be produced by the collapse of rapidly rotating massive stars \citep[collapsars,][]{1993ApJ...405..273W, 1999ApJ...524..262M, 2006ARA&A..44..507W}, and in this situation a long-lasting X-ray emitting jet might be produced by continuous fallback of an extended stellar envelope \citep{kumar08_longGRB_disk_evolution}, although it is unclear whether the energy injection from the disk wind and jet during the prompt emission phase could completely unbind the outer stellar envelope \citep{2010ApJ...713..800L, gottlieb22_MHDjet_cocoon}. For short GRBs from NS-NS or NS-BH mergers, it has been proposed that a fraction of the bound gas can fall back on longer timescales and fuel the accretion \citep[e.g.,][]{rosswog07_fallback, lee09_disk_phase_transition, cannizzo11_fallback_disk, gibson17_fallback_accretion}. Along this line of thought, \citet{kisaka15_fallback_power} propose that the magnetic field topology near the black hole (BH) evolves due to interactions with the fallback material in a way such that the \citet{blandford77_BZjet} jet power stays constant with time during the EE phase. However, mechanical feedback from the disk wind/jet \citep{fernandez15_wind_fallback_interaction} as well as radioactive heating by r-process nuclei \citep{metzger10_heating_fallback, desai19_fallback_heating, ishizaki21_fallback_heating} can suppress\footnote{\citet{desai19_fallback_heating} argue that fallback accretion producing EE is likely restricted to NS-BH systems, while in this work we propose that NS-NS mergers could have long-lived accretion disks and hence could power EE. This may be tested by future GW data.} the fallback rate of marginally bound material (those with fallback time longer than $\sim$1 second). More detailed calculations on the hydrodynamic interactions between the marginally bound material and existing accretion disk are needed to reliably test the viability of the fallback accretion model (especially to determine if the model can produce a sharp X-ray flux cutoff at $t\sim 10^2\rm\,s$).


Another popular model for the EE and X-ray plateau is based on the dipole spindown power of a millisecond magnetar --- strongly magnetized NS \citep[e.g.,][]{metzger08_magnetar_EE, metzger11_magnetar_GRB}. One scenario to explain the steep flux decline is that the central object needs is a supramassive NS temporarily supported by rigid-body rotation until $t\sim10^2\rm\, s$ when it collapses into a BH \citep{2014ApJ...785...74L, 2015ApJ...805...89L, 2016PhRvD..93d4065G}. In this scenario, there is no physical explanation why the collapse to a BH due to angular momentum loss by GW and magnetic torques would preferentially occur on a timescale of $10^2\rm\, s$ in most of the observed systems. In fact, if a supramassive NS is indeed formed in a large fraction of NS mergers, a very broad distribution of the collapse time is expected \citep{beniamini21_SMNS_collapse_time}. 
Within the magnetar framework, another scenario to explain the steep flux decline at the end of the EE/plateau is that the magnetization of NS wind rapidly increases as the neutrino-driven mass loss rate from the NS drops on a timescale of $\sim 10^2\rm\, s$.

\begin{figure*}
  \centering
\includegraphics[width = 0.48\textwidth]{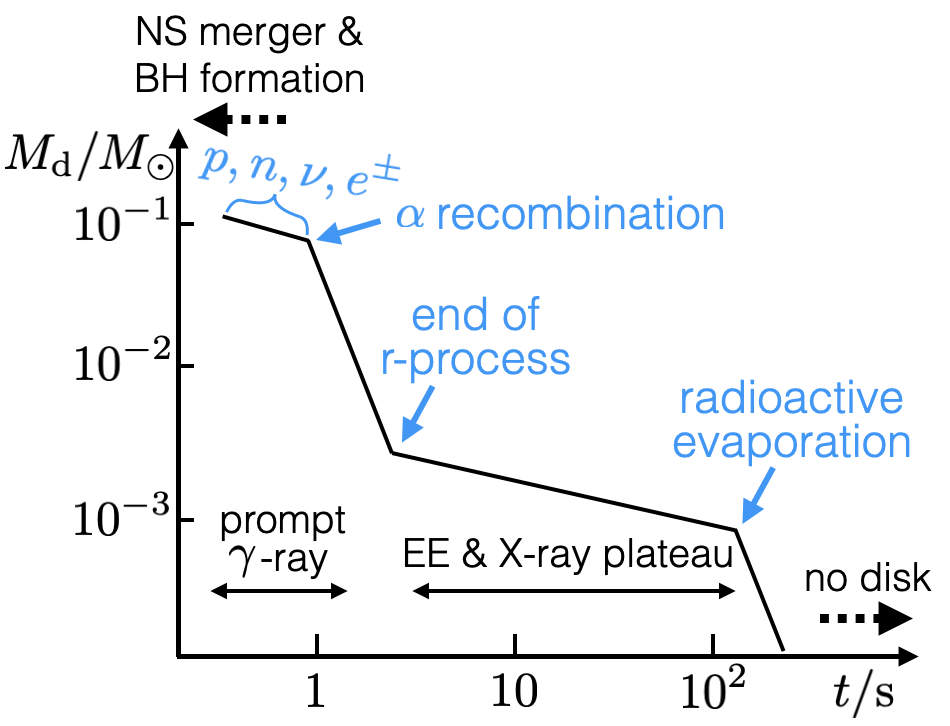}
\includegraphics[width = 0.4\textwidth]{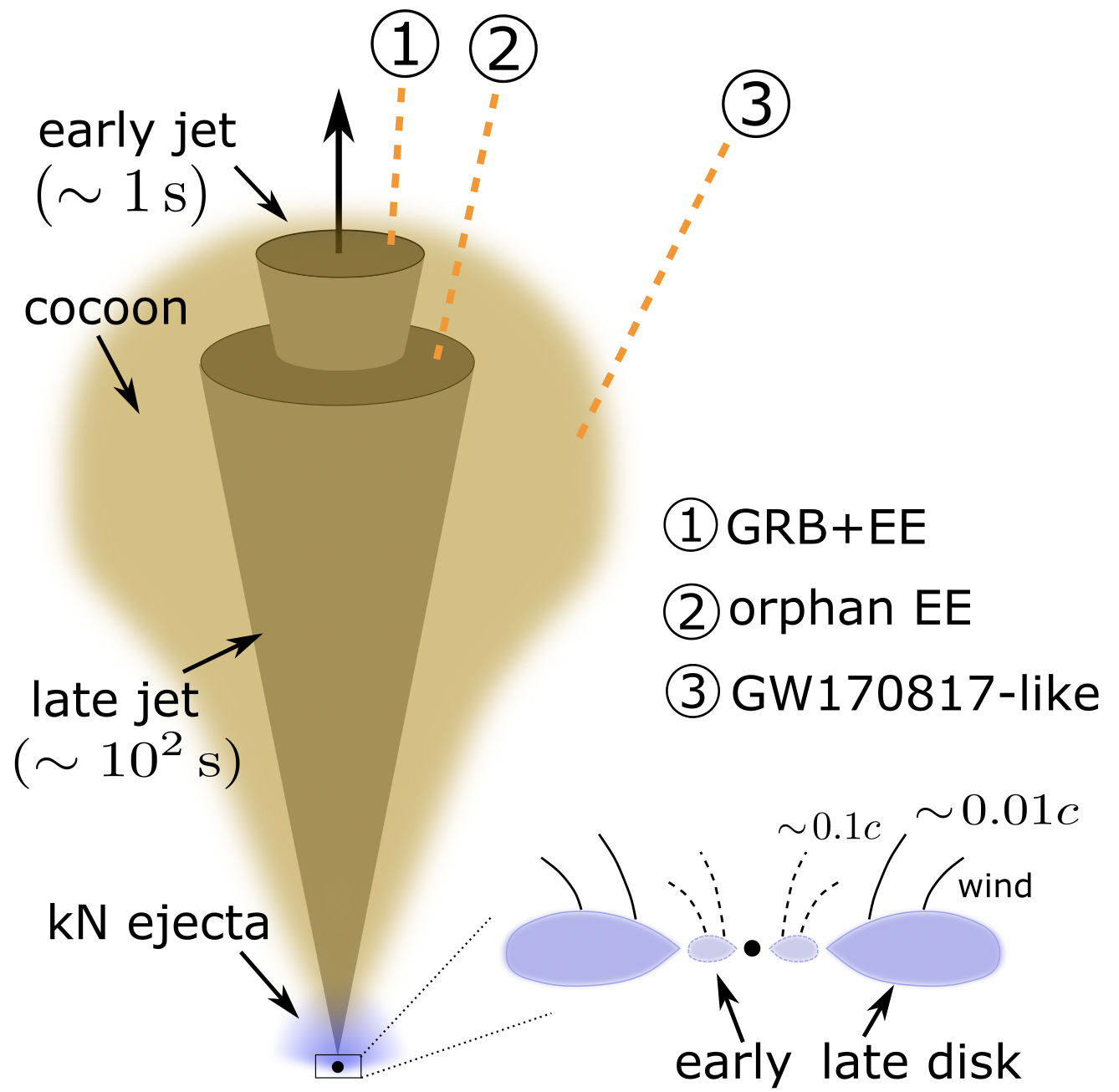}
\caption{\textit{Left panel}: A schematic picture for the time evolution of the accretion disk mass after a NS-NS merger. In this picture, the merger remnant collapses into a BH at early time ($t < 1\rm \,s$), and the neutrino-dominated accretion disk (consisting of free nucleons, neutrinos, pairs, and magnetic fields) powers an energetic jet that produces the prompt $\gamma$-ray emission in the first $\mathcal{O}(1)\rm\, s$. As the disk viscously spreads and cools, nuclear recombination leads to the formation of $\alpha$ particles and subsequently r-process elements. The release of nuclear binding energy drives the majority of the disk mass unbound, and this violent mass loss phase ends when r-process nucleosynthesis is finished (corresponding to an abrupt drop in the heating rate, see Fig. \ref{fig:heating_rates}). Subsequently, the leftover disk continues to accrete onto the BH, which powers a long-lived jet that produces the EE and X-ray plateau. Finally, when the cumulative radioactive heating by r-process nuclei exceeds the binding energy of the viscously spreading material at $t\sim \mathcal{O}(10^2)\rm\, s$, the entire disk rapidly evaporates, which corresponds to the steep decline in the observed $\gamma/$X-ray flux. Eventually, no disk is left behind.\\
\textit{Right panel}: A broad-brush picture of a two-component jet, which is a natural consequence of our disk evolution model for NS mergers. In \S \ref{sec:discussion}, we present evidence supporting a scenario in which the jet component from late-time accretion (corresponding to the EE) has a wider beaming angle than that of the prompt $\gamma$-ray jet component. A physical motivation for $\Omega_{\rm EE}>\Omega_\gamma$ is that the early jet is strongly collimated by a high-pressure cocoon due to interactions with pre-existing gas near the poles whereas the late jet propagates in the cavity created by the laterally expanding cocoon. In this case, at different viewing angles, the observer would see different high-energy emission, as illustrated by the three cases: \circled{1} detecting both prompt $\gamma$-rays and EE, \circled{2} EE without prompt $\gamma$-rays, \circled{3} only emission from the cocoon surrounding the jet. The model is testable because the afterglow emission for these viewing angles are also different. Another prediction of the model is that the late-time disk wind (due to radioactive evaporation) generates narrow line features in the kilonova (kN) spectrum with $\Delta v \sim 0.01c$, which are detectable by JWST a few weeks after the merger.
}\label{fig:sketch}
\end{figure*}

A potential issue of any NS-spindown-powered scenario is that the large amount ($\gtrsim10^{52}\rm\, erg$) of energy injection into the surrounding medium would produce exceedingly bright radio emission which is severely constrained by increasingly stringent upper limits \citep[e.g.,][]{shroeder20_radio_upper_limits, bruni21_radio_upper_limits}. Another potential issue is that if the spindown energy is used to generate the EE/plateau, then the prompt emission (which has much higher luminosity and shorter duration) still requires an ultra-relativistic jet driven by hyper-accretion (of $0.01$--$0.1\msun$) onto the NS, which has so far not been numerically demonstrated. From the angular momentum transport inside the merger remnant (which is, for most cases, initially a hypermassive NS supported by differential rotation), most NS mergers likely form a BH within the first second \citep{margalit22_remnant_fate}.

In this work, we explore an alternative model for the late-time central engine activities in GRBs based on long-lived disk accretion, under the assumption that a fraction of the disk mass survives the violent outflow phase of nuclear recombination. In the most energetically efficient scenario, only a small disk mass ($>10^{-4}\msun$) is needed to power the observed EE and X-ray plateau. The more challenging features to explain are the duration of the EE/X-ray plateau and the shallow flux decay before the steep decline. In our model, the jet shuts off when the disk evaporates as a result of the cumulative radioactive heat exceeding the binding energy at late time, and the shallow flux decay is because jet power is linearly proportional to the disk mass (but NOT the accretion rate). A schematic picture of our model is shown in Fig. \ref{fig:sketch}.

If this model is true, the late-time disk evaporation provides slower ejecta ($v\sim 0.01c$) which will produce narrow line features in the kilonova spectra observable at $t\gtrsim 20\rm\, d$. Potential detections of such narrow lines by \textit{James Webb Space Telescope} (JWST) provides another motivation for this work.




This paper is orgainzed as follows. In \S \ref{sec:disk_model}, we calculate the disk evolution under two possible scenarios (\S\ref{sec:without_wind} and \S\ref{sec:with_wind}) that differ on the prescriptions of MHD winds from advection-dominated accretion flows. In \S \ref{sec:jet}, we estimate the jet power based on the \citet{blandford77_BZjet} formalism assuming the remnant is a rapidly spinning BH and compare it with GRB observations. In \S \ref{sec:kilonova}, we estimate the contribution to late-time kilonova emission by the slow ejecta and discuss JWST detectability. We discuss observational constraints on our model in \S \ref{sec:discussion}. A summary is provided in \S \ref{sec:summary}.

\section{Model}\label{sec:disk_model}
We assume that the merger remnant has collapsed to a black hole (BH) early on (at $t< 1\rm\, s$ after the merger) and that it only affects the outer disk evolution through Newtonian gravity. Our fiducial value for the remnant mass is $M=2.6\msun$, as expected from the mass distribution of the Galactic double NS population after accounting for the mass loss due to GW, neutrinos and baryonic ejecta \citep{farrow19_Galactic_BNS_masses}. We approximate the radial distribution of the disk mass as a ``ring'' located at the disk outer radius $\rd$ where the surface density distribution peaks, and calculate the time evolution of the thermodynamic quantities near the mid-plane at radius $\rd$ --- pressure $P$, density $\rho$, temperature $T$, etc.

As a result of angular momentum transport by viscous stresses, the disk mass $M_{\rm d}$ and radius $\rd$ evolve as
\begin{equation}
    \label{eq:evolution}
    \dotMd = -\Md/\tvis,\ \dot{r}_{\rm d} = {2\rd/ \tvis},
\end{equation}
where the viscous timescale is given by
\begin{equation}
  \label{eq:tvis}
  t_{\rm vis} = \rd^2/\nu = (\alpha \theta^2 \OmgK)^{-1},
\end{equation}
$\Omega_{\rm K} = \sqrt{GM/\rd^3}$ is the Keplerian angular frequency, $\nu=\alpha c_{\rm s}H$ is the kinetic viscosity \citep{1973A&A....24..337S} most likely due to MHD turbulence \citep{balbus98_MRI_turbulence}, $c_{\rm s} = \sqrt{P/\rho} = H\Omega_{\rm K}$ is the isothermal sound speed, $H$ is the disk scaleheight, and $\theta=H/\rd$ is the dimensionless scaleheight.

The mass evolution in eq. (\ref{eq:evolution}) is due to accretion towards smaller radii as well as unbound outflow driven by viscous heating \citep{yuan14_ADAF_review} --- the fractional contributions from these two components are uncertain but our results are qualitatively insensitive to which dominates $\Md/\tvis$. The radius evolution in eq. (\ref{eq:evolution}) is due to angular momentum conservation and the expression roughly holds even when a large fraction of the disk mass loss $\dotMd$ is due to wind, provided that the wind has the same specific angular momentum as that in the disk (i.e., the wind does not provide an additional torque on the remaining gas in the disk). In the case where the wind carries less (more) specific angular momentum, the disk expands faster (slower) than in our model and hence the disk mass and accretion rate drop faster (slower) \citep[see e.g.,][]{kumar08_longGRB_disk_evolution}. 

For given initial conditions $\rd(t=t_0)=r_{\rm d,0}$ and $\Md(t=t_0)=M_{\rm d,0}$, one can analytically integrate eq. (\ref{eq:evolution}) using the viscous time in eq. (\ref{eq:tvis}), under the assumption that the dimensionless disk height $\theta$ stays constant. The well-known result is \citep[see e.g.,][]{metzger08_onezone_model}
\begin{equation}\label{eq:self_similar}
\begin{split}
    \Md(t) &= M_{\rm d,0} \lrsb{1 + 3(t-t_0)/t_{\rm vis,0}}^{-1/3},\\
    \rd &= r_{\rm d,0} \lrsb{1 + 3(t-t_0)/t_{\rm vis,0}}^{2/3},\\
    \tvis &= 3(t-t_0) + t_{\rm vis,0},
\end{split}
\end{equation}
where $t_{\rm vis,0} = \alpha^{-1}\theta^{-2}\sqrt{r_{\rm d,0}^3/GM}$ is the initial viscous time. We emphasize that $M_{\rm d,0}$ is the disk mass at the end of nuclear recombination (but NOT right after the merger). The energy release due to nuclear recombination is capable of driving the majority of the initial disk mass unbound in the first few seconds, so we expect $M_{\rm d,0}\ll 0.1\msun$. The global GRMHD simulations by \cite{siegel18_GRMHD_disk, fernandez19_long_GRMHD} show that roughly 1\%--10\% of the initial disk mass stays bound at the end helium recombination, although there is so far no self-consistent, long-term simulation that includes dynamically coupled nucleosynthesis beyond He with MHD accretion. Taking their results at face value, and for an initial torus mass of the order $0.1\msun$, we expect $M_{\rm d,0}$ to be in the range $10^{-3}$--$10^{-2}\msun$.

In reality, the disk height at a given time is determined by the vertical pressure gradient which in turn is determined by the balance between heating and cooling rates. Since the disk material achieves local thermodynamic equilibrium in the vertical direction on a sound-crossing timescale which is much shorter than the viscous time, the energy conservation equation can be written as
\begin{equation}
  \label{eq:energy}
  q_{\rm vis} + q_{\rm h} = q_{\rm adv} + q_{\rm w},
\end{equation}
where the heating terms on the LHS are the viscous heating rate per unit mass $q_{\rm vis}$ and radioactive heating rate $q_{\rm h}$, and the cooling terms on the RHS are the rate at which heat is advected towards smaller radii by the mass inflow $q_{\rm adv}$ and the wind cooling rate $q_{\rm w}$ which is important if a large fraction of $\dotMd=-\Md/\tvis$ is due to viscously driven wind. We note that, due to our incomplete understanding of geometrically thick or advection-dominated disks \citep[see][for a review]{yuan14_ADAF_review}, the wind cooling term $q_{\rm w}$ is rather uncertain, and an attempt to treat this term is presented in \S\ref{sec:with_wind} where we demonstrate that our qualitative result (that the disk rapidly evaporates at late time due to radioactive heating) is robust to uncertainties in the wind prescription.


In the following we estimate each of the heating/cooling rates. Due to radial pressure gradient, the disk rotates at a sub-Keplerian frequency roughly given by
\begin{equation}
  \label{eq:omega}
  \Omega \simeq {\OmgK\over 1 + \theta^2}.
\end{equation}
Viscous heating occurs as a result of frictional angular momentum exchange between adjacent annuli and the heating rate is given by
\begin{equation}
\label{eq:qvis}
    q_{\rm vis} \simeq {9\over 4} \nu \Omega^2 \simeq {9\over 4} {GM/\rd \over (1+\theta^2)^2 \tvis}.
\end{equation}
The radial mass inflow advects heat to smaller radii \citep{narayan94_adaf1} and hence contributes to an advective cooling rate approximately given by
\begin{equation}
\label{eq:qadv}
  q_{\rm adv} = v_{\rm r}T{\partial s\over \partial r} \simeq
  {3\over 2} \lrb{{U\over P} - 1} \theta^2 {GM/\rd \over t_{\rm vis}},
\end{equation}
where $s$ is the specific entropy, and $v_{\rm r} = -3\nu_{\rm vis}/(2\rd)$ is the radial velocity of the mass inflow. To obtain the second expression in eq. (\ref{eq:qadv}), we have made use of $T\d s = \d h - \d P/\rho$, $h=(U+P)/\rho\propto c_{\rm s} \propto r^{-1}$ being the specific enthalpy ($U$ being the thermal energy density) and $\partial h/\partial r\simeq -h/\rd$, and we have taken an approximate radial pressure scaling of $P\propto r^{-2}$ (and hence $\partial P/\partial r\simeq -2P/\rd$) in between extreme cases of no wind loss ($P\propto r^{-2.5}$) and strongest wind loss ($P\propto r^{-1.5}$) \citep{blandford99_ADIOS}. We adopt the $\mathtt{Helmholtz}$\footnote{We used a convenient Python implementation provided by M. Coleman at \href{https://github.com/msbc/helmeos}{https://github.com/msbc/helmeos}.} \citep{timmes00_helmeos} equation of state $P(\rho, T)$ and $U(\rho, T)$ for a mean atomic mass number $\bar{A}=100$ and mean charge number $\bar{Z}=0.4\bar{A}$. For the parameter space in this work, the pressure is dominated by electrons, positrons and radiation while the contribution from ions is negligible, so our results are practically unaffected by the choice of $\bar{A}$ and $\bar{Z}$.

Finally, the radioactive heating is due to the decay of the freshly synthesized r-process elements inside the disk. For an initial $\Ye\lesssim 0.4$, the heating rates after the end of r-process in the time range $10\lesssim t\lesssim 10^4\rm\, s$ are insensitive to the initial conditions \citep[e.g.,][]{metzger10_kilonova} and can be approximated by the following power-law (see Appendix \S \ref{sec:heating_rate})
\begin{equation}
\label{eq:qh}
    q_{\rm h}\approx 4\times10^{16} (t/\mr{s})^{-1.3}\rm\, erg\,s^{-1}\,g^{-1},
\end{equation}
where we have subtracted 30\% of the total heating rate due to neutrinos accompanied by $\beta$-decays \citep{barnes16_heating_thermalization}.

As the disk viscously expands following the scaling of $\rd\propto t^{2/3}$, the viscous heating rate has temporal scaling $q_{\rm vis}\propto t^{-5/3}$. The advective and wind cooling rates have the same scaling as the viscous heating rate. Thus, at sufficiently late time, the radioactive heating rate must exceed the viscous heating rate, and the solution to eq. (\ref{eq:energy}) will be $\theta\gtrsim 1$ --- which is unphysical. We propose that the excessive heating will rapidly evaporate all the disk material. When the disk material becomes unbound, we modify the mass conservation equation in eq. (\ref{eq:evolution}) by adding an evaporation term $\dot{M}_{\rm evap}=-\Md\OmgK$ to capture rapid mass removal, where $\OmgK^{-1}$ is the sound-crossing time in the vertical direction. In the following, we try two different approaches to quantify the ``boundness'' of the disk material --- the first approach ignores wind cooling term $q_{\rm w}$ in the energy conservation equation and the second takes into account the possibility that the disk is cooled by a viscously driven wind  --- and the results are qualitatively similar.

We note here a limitation of our simplified one-zone disk model. When the outer disk starts to evaporate as a result of excessive heating, the inner disk material at smaller radii $r\ll \rd$ can remain bound despite the loss of pressure confinement and will evaporate later. Thus, the rate at which the total disk mass evaporates with time will be somewhat slower than $\dot{M}_{\rm evap}=-\Md\OmgK$.
The self-consistent solution can only be captured by higher dimensional hydrodynamic simulations of the system, which are left for future works.

\subsection{Sound Speed-limited Disk without Wind}\label{sec:without_wind}

In this subsection, we ignore the wind cooling term by setting $q_{\rm w}=0$ and this allows the scaleheight of a disk without radioactive heating (in the limit $q_{\rm h}\ll q_{\rm vis}$) to be moderately high $\theta\sim 0.5$. A well-known consequence is that the Bernoulli number (see eq. \ref{eq:Bernoulli_number}) of the fluid elements in the outer disk is positive \citep{narayan94_adaf1, blandford99_ADIOS}. In this scenario, we impose a mass evaporation term when the sound speed $\cs$ exceeds the orbital speed $\Omega\rd$, and the mass conservation equation becomes
\begin{equation}
\label{eq:evolution_mass_cs}
    \dotMd = -{\Md/\tvis} - \Md\OmgK\, \mr{Sig}\lrb{\cs - \Omega\rd \over \Omega \rd},
\end{equation}
where, instead of an abrupt transition at $\cs=\Omega \rd$, we use a Sigmoid function to achieve numerical smoothness,
\begin{equation}
\label{eq:sigmoid}
    \mr{Sig}(x) = {1\over 1 + \exp(-x/\delta)},
\end{equation}
and $\delta=1/300$ describes the sharpness of the transition.

\begin{figure}
  \centering
\includegraphics[width = 0.48\textwidth]{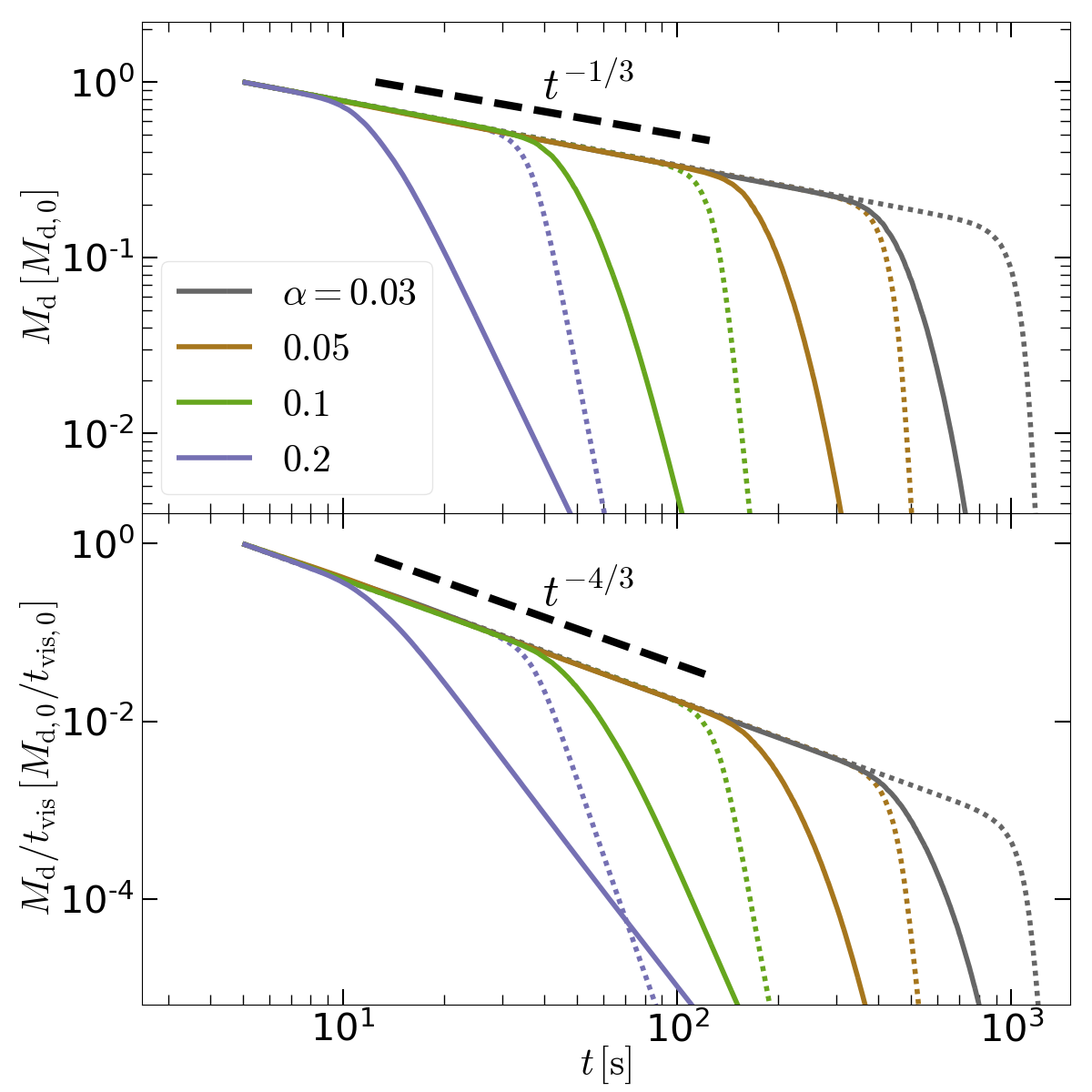}
\caption{Temporal evolution of the disk mass $\Md$ and accretion rate $\Md/\tvis$ in different models. We normalize the disk mass by its initial value $M_{\rm d,0}$ after heavy nuclei formation at $t_0=5\rm\, s$ and the accretion rate by the initial parameters $M_{\rm d,0}/t_{\rm vis,0}$. Solid lines are for the sound speed-limited windless disk model (\S\ref{sec:without_wind}), whereas dotted lines are for Bernoulli-limited disk with wind cooling and for $B_{\rm e,0}=-0.1$ (\S\ref{sec:with_wind}). Different colors represent different viscous parameter $\alpha$, ranging from 0.03 (grey) to 0.2 (purple). The BH mass is fixed at $M=2.6\msun$. We find that, for $\alpha=0.1$, the disk rapidly evaporates (due to excessive radioactive heating) roughly $10^2\rm\, s$ after the merger. A larger $\alpha$ corresponds to earlier evaporation.
}\label{fig:disk_evolution}
\end{figure}

We then numerically integrate the disk radius evolution in eq. (\ref{eq:evolution}) and mass evolution in eq. (\ref{eq:evolution_mass_cs}) (with the disk temperature given by eq. \ref{eq:energy}), for an initial time $t_0=5\rm\, s$ after the merger (at the end of the nucleosynthesis) and initial disk radius such that $t_{\rm vis,0}(r_{\rm d,0}) = 3t_0$. The late-time evolution at $t\gg t_0$ depends weakly on the choices of initial conditions. For our fiducial case, the initial disk mass is taken to be $M_{\rm d,0}=3\times 10^{-3}\msun$, and the disk evaporation time depends weakly on the choice of $M_{\rm d,0}$ (which affects the equation of state through disk density). The results of the disk evolution are shown by the solid lines in Fig. \ref{fig:disk_evolution}, for a number of choices of viscous parameter $\alpha$ in the range between 0.03 and 0.2.

\begin{figure}
  \centering
\includegraphics[width = 0.48\textwidth]{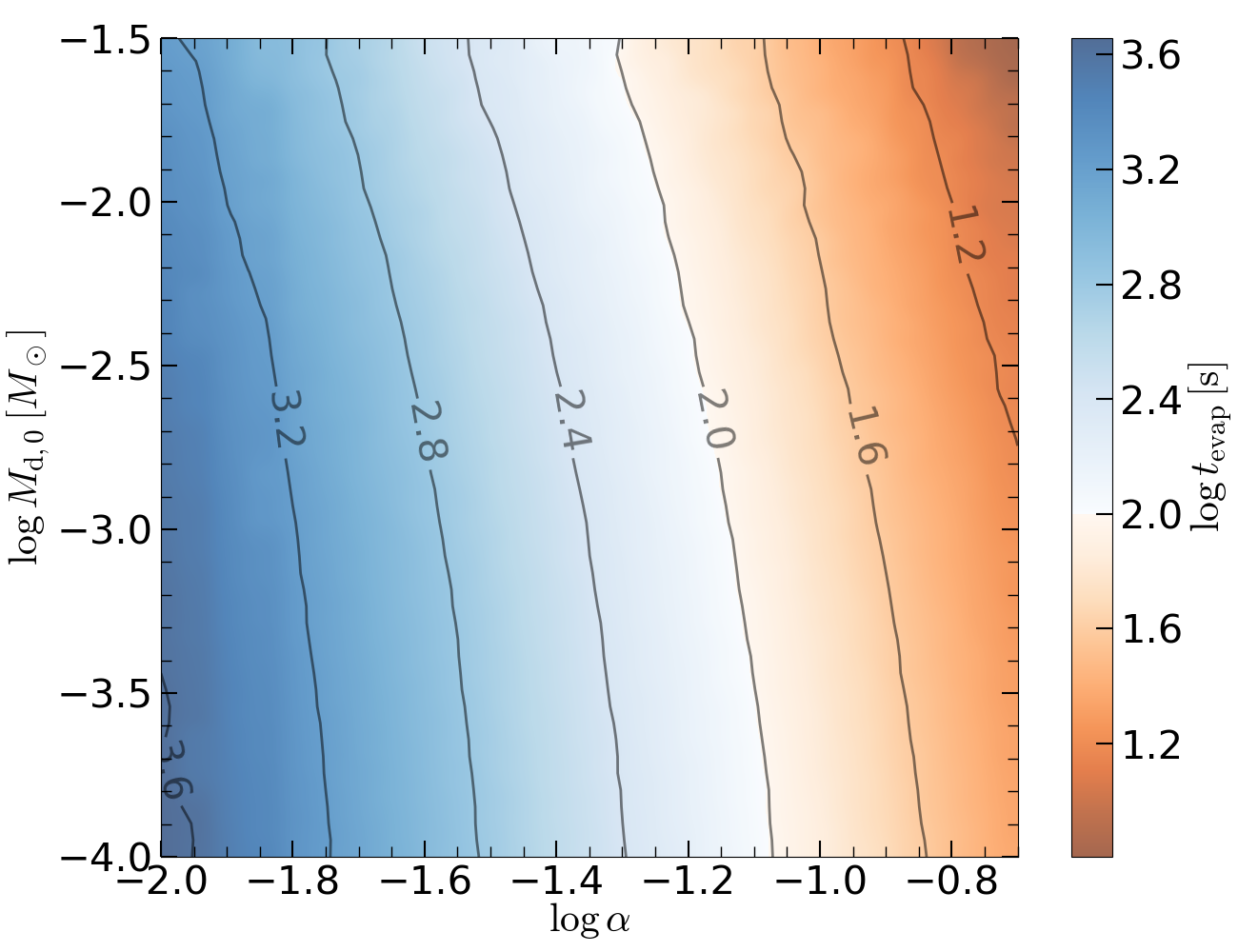}
\caption{Disk evaporation time $\tevap$ (defined as when $\d\log\Md/\d\log t$ first drops below $-3$) as a function of the disk mass $M_{\rm d,0}$ at the end of nuclear formation and viscous parameter $\alpha$, for the windless evolution model in \S\ref{sec:without_wind}. The BH mass is fixed at $M=2.6\msun$.
}\label{fig:tevap_nofw}
\end{figure}

We find that the disk mass initially evolves slowly with time $\Md\propto t^{-1/3}$ until a critical time when the disk sound speed exceeds its rotational speed, and then, according to our model prescription, the disk rapidly evaporates on a sound-crossing timescale of $\OmgK^{-1}$. The disk accretion rate initially evolves as $\dotMd\propto t^{-4/3}$ and then quickly drops after the critical time for this phase transition. We define the evaporation time $\tevap$ as when $\d\log\Md/\d\log t$ first drops below $-3$ (as a signature of rapid X-ray flux drop). This depends strongly on the viscous parameter but weakly on the initial disk mass $M_{\rm d,0}$, as shown in Fig. \ref{fig:tevap_nofw}. Since disk evaporation occurs at roughly a fixed ratio between the cumulative heating and the disk binding energy $q_{\rm h} t/(GM/\rd)$, we obtain $\tevap^{-0.3}\rd(\tevap)/M\simeq \rm const.$ and hence the following scaling
\begin{equation}
    \tevap\simeq 50\mr{\,s}\, \lrb{\alpha\over 0.1}^{-1.8}\lrb{M\over 2.6\msun}^{1.8},
\end{equation}
where we have used $\rd\propto t^{2/3} M^{1/3} \alpha^{2/3}$ (from eq. \ref{eq:tvis}) and the normalization is taken from the $M_{\rm d,0}=3\times 10^{-3}\msun$ case. Note that the dependence of $\tevap$ on the BH mass $M$ is not shown in Fig. \ref{fig:tevap_nofw}.


\subsection{Bernoulli-limited Disk with Wind}\label{sec:with_wind}

In this subsection, we describe another scenario where a fraction of the viscously driven mass evolution $\Md/\tvis$ is taken away by the disk wind and the wind tries to cool the remaining disk material such that its Bernoulli number is maintained at a negative or energetically bound value $B_{\rm e,0}<0$  \citep[see][for a similar treatment but in a different context]{margalit16_NSWD_merger}. The dimensionless Bernoulli number is defined as
\begin{equation}
  \label{eq:Bernoulli_number}
  B_{\rm e} = {(U+P)/\rho + \Omega^2\rd^2/2 - GM/\rd \over GM/\rd} \approx {U\over P} \theta^2 - 0.5,
\end{equation}
where we have ignored a small kinetic energy term associated with the radial velocity of the disk material. We consider $B_{\rm e,0}=-0.1$ as a fiducial value, which is motivated by recent numerical simulations of adiabatic accretion flows \citep[][their ``most realistic'' Model B]{yuan12_Be_ADAF} \citep[see also the ``SANE'' model of][]{narayan12_ADAF}. We note that the initial conditions of these simulations are not necessarily realistic as in NS mergers, and more work (with larger numerical domains and longer runs with more realistic magnetic fields) is needed to get a more reliable $B_{\rm e,0}$. The overall expectation is that the closer $B_{\rm e,0}$ is to zero, the easier it is for extra radioactive heating to unbind the disk and the sooner the disk evaporates.

We assume the wind cooling rate to be in the following form
\begin{equation}
\label{eq:qw}
    q_{\rm w} = f_{\rm w} {GM/\rd \over \tvis},
\end{equation}
where $f_{\rm w}$ is a strength parameter of order unity that depends on the detailed wind mass loss rate fraction (out of $\Md/\tvis$) and the asymptotic specific energy of the wind. To maintain $B_{\rm e}=B_{\rm e,0}$, the wind cooling strength parameter for a disk without radioactive heating (in the limit $q_{\rm h}\ll q_{\rm vis}$) can be found to be
\begin{equation}
\label{eq:fw}
    f_{\rm w} = {9\over 4} {1\over (1+\theta_0^2)^2} - {3\over 2}\lrb{{U\over P} - 1}\theta_0^2,
\end{equation}
where $\theta_0$ is the characteristic disk height given by (from eq. \ref{eq:Bernoulli_number})
\begin{equation}\label{eq:Be0}
    \theta_0^2 = (B_{\rm e,0} + 0.5) P/U.
\end{equation}
For our fiducial choice of $B_{\rm e,0}=-0.1$, the characteristic disk height is $\theta_0\simeq 0.4$, which is slightly smaller than that in the $q_{\rm w}=0$ case. 

One can plug the wind cooling rate given by eqs. (\ref{eq:qw}) and (\ref{eq:fw}) into the energy conservation equation (\ref{eq:energy}) to solve for the disk height $\theta$ at a given time. At sufficiently late time, the Bernoulli number must become positive as a result of excessive radioactive heating and we still expect the disk to evaporate on a sound-crossing time. Thus, we impose a mass evaporation term and the mass conservation equation becomes
\begin{equation}
    \label{eq:evolution_mass_Be}
    \dotMd = -{\Md/\tvis} - \Md\OmgK\, \mr{Sig}\lrb{B_{\rm e}},
\end{equation}
where the Sigmoid function is given by eq. (\ref{eq:sigmoid}).

We then numerically integrate the disk radius evolution in eq. (\ref{eq:evolution}) and mass evolution in eq. (\ref{eq:evolution_mass_Be}), for an initial time $t_0=5\rm\, s$ since the merger and initial disk radius such that $t_{\rm vis,0} = 3t_0$ (same as in \S\ref{sec:without_wind}). We take a fiducial initial disk mass of $M_{\rm d,0}=3\times 10^{-3}\msun$, which only weakly affects the disk evaporation time. The results of the disk evolution are shown by the dotted lines in Fig. \ref{fig:disk_evolution}. For the case of $B_{\rm e,0}=-0.1$, the overall disk evolution is qualitatively similar to the windless case with $q_{\rm w}=0$ in \S \ref{sec:without_wind}. The disk evaporation occurs slightly later than in the windless model, but the time of this transition depends on our choice of $B_{\rm e,0}$.

\begin{figure}
  \centering
\includegraphics[width = 0.48\textwidth]{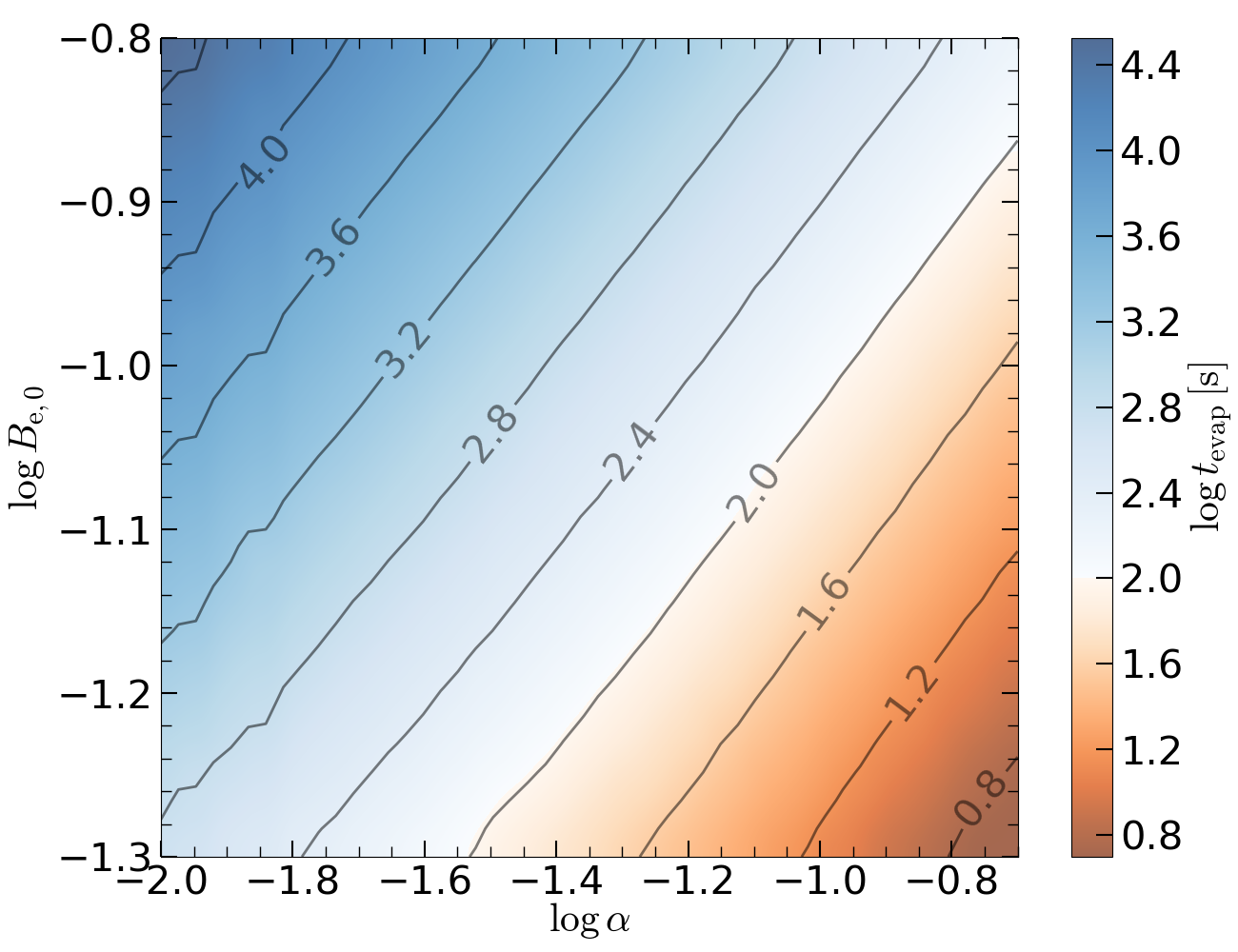}
\caption{Disk evaporation time $\tevap$ (defined as when $\d\log\Md/\d\log t$ first drops below $-3$) for the Bernoulli-limited disk evolution including wind cooling in \S \ref{sec:with_wind}. The BH mass is fixed at $M=2.6\msun$.
}\label{fig:tevap_fw}
\end{figure}

The evaporation time $\tevap$, as defined by the earliest time when $\d\log\Md/\d\log t=-3$ (same as in \S\ref{sec:without_wind}), as a function of $\alpha$ and $B_{\rm e,0}$ is shown in Fig. \ref{fig:tevap_fw}. Since disk evaporation occurs roughly at a fixed ratio of $q_{\rm h} t/(|B_{\rm e,0}|GM/\rd)$, we obtain $\tevap^{-0.3}\rd(\tevap)/(|B_{\rm e,0}|M)\simeq \rm const.$ and hence the following scaling
\begin{equation}
    \tevap\simeq 100\mr{\,s}\, \lrb{\alpha\over 0.1}^{-1.8}\lrb{M\over 2.6\msun}^{1.8} \lrb{|B_{\rm e,0}|\over 0.1}^{2.7},
\end{equation}
where the normalization approximately reproduces the numerical values in Fig. \ref{fig:tevap_fw}.

We note that $\tevap$ strongly depends on the critical Bernoulli number $B_{\rm e,0}$ in our model. This means that the model is not strongly predictive without a careful calibration against numerical simulations which can provide a better physical understanding of the viscously driven disk wind (as parameterized by $f_{\rm w}$ in the current model). Nevertheless, our very simple model is able to reproduce the duration of the EE and X-ray plateau (or the time of the rapid X-ray flux drop) with reasonable parameters of $\alpha\simeq 0.1$ and $B_{\rm e,0}\simeq -0.1$ that are consistent with existing global MHD simulations of adiabatic accretion flows \citep{narayan12_ADAF, liska20_disk_dynamo} \citep[however, some other simulations predict lower $\alpha$, e.g.,][]{davis10_alpha_shearbox, hawley11_alpha_GRMHD}. Observations of fully ionized geometrically thin (i.e., radiatively efficient) disks indicate $\alpha\sim 0.1$ but it is unclear if this applies to thick disks. Better calibration of our model can be obtained by performing GRMHD simulations with more realistic initial conditions (better corresponding to NS mergers) and longer runs lasting for many viscous timescales of the outer disk.

A final note is that since the evaporation time scales with the BH mass as $\tevap\propto M^{1.8}$, it is likely that the steep flux decline at the end of long GRB X-ray plateaus (at $t\sim 10^4\rm\, s$) is simply due to more massive BHs in those cases.

\section{Jet Power}\label{sec:jet}

The properties in the outer disk provides the boundary conditions for the inner regions of the disk, which extends from $\rd$ all the way down to the horizon. The inner boundary condition, for the case of a BH remnant, is provided by freefall at the event horizon. In this section, we assume that the jet is powered by the Poynting flux associated with the rotating B-field lines that are frame-dragged by the BH spin  \citep{blandford77_BZjet}. It should be noted that the physical processes related to GRB jet launching are still poorly understood, so the model to be presented in this subsection is only a possibility that is not ruled out by observations \citep[see][for a review of other possibilities]{kumarzhang15_GRB_review}.



In the \citet{blandford77_BZjet} picture, the jet power is roughly given by \citep{tchekhovskoy10_BZpower}
\begin{equation}
    L_{\rm BZ} \simeq {c\over 96\pi^2 \rg^2} \PhiBH^2 \omega_{\rm BH}^2,
\end{equation}
where $\PhiBH$ is the (poloidal) magnetic flux accumulated on the BH horizon and $\omega_{\rm BH}$ is the angular frequency of the horizon due to frame dragging
\begin{equation}
    \omega_{\rm BH} = {a\over 1 + \sqrt{1-a^2}}.
\end{equation}
For nearly equal mass NS binaries, the typical spin of the merger remnant is $a\sim 0.7$ \citep{gonzalez07_merger_kick_spin, kiuchi09_BH_remnant, rezzolla10_torus_mass, sekiguchi16_BH_remnant, dietrich17_bNS_ejecta_mass}, provided that a large fracton the angular momentum of the binary system is retained by the remnant BH (although this depends on the NS equation of state and the uncertain evolution of the merger remnant). Thus, we take $\omega_{\rm BH}=0.4$ as a fiducial value.


After the merger, the magnetic field strength is expected to be dominated by the toroidal component, and it has been shown that turbulent motions in the outer disk may generate a large-scale poloidal component in a process similar to the $\alpha\omega$-dynamo \citep{liska20_disk_dynamo}. Motivated by this idea, we assume that, when the system reaches a saturated state, the squared ratio between the poloidal B-field component and the total (toroidal-dominated) B-field is
\begin{equation}
\Bp^2/B^2 = \xi.
\end{equation}
Note that here $\Bp$ represents the volume- and time-averaged net poloidal field that is coherent on a lengthscale of the disk scaleheight $H=\theta \rd$, so the total poloidal flux in the outer disk is given by
\begin{equation}
    \Phid \simeq \Bp H^2 = \xi^{1/2}\theta^2 B \rd^2 = \xi^{1/2} \theta^2 \lrb{P_{\rm B}\over P}^{1/2} \sqrt{8\pi P} \rd^2,
\end{equation}
where $P_{\rm B} = B^2/8\pi$ is the total magnetic pressure. For a radiation-dominated system in virial equilibrium, the total pressure can be estimated by $P\simeq U/3\sim \rho (GM/2\rd)/3$. Since the density in the outer disk is $\rho\simeq \Md/(2\pi \rd^2 H)$, we obtain the total poloidal flux
\begin{equation}
    \Phid \simeq \xi^{1/2} \theta^{3/2} \lrb{P_{\rm B}\over P}^{1/2} \sqrt{2GM\Md/3}.
\end{equation}
Over a viscous timescale, a significant fraction of the poloidal flux in the outer disk may accumulate onto the BH horizon as a result of accretion, so one may expect that the magnetic flux on the horizon to also scale as $\PhiBH \propto \Md^{1/2}$, provided that the ratio $\PhiBH/\Phid$ is roughly a constant of order unity. In reality, $\PhiBH$ is the net flux accumulated over the accretion history up to the current time. For toroidal-dominated fields in the disk, we expect the poloidal fields arriving at the horizon to undergo  cancellations such that the long-term averaged poloidal flux is negligible whereas the instantaneous value roughly tracks $\Phid$ in the outer disk.

Based on the above arguments, the jet power can be written in the following form
\begin{equation}\label{eq:LBZ}
\begin{split}
    L_{\rm BZ}\simeq &\, 5\times10^{48}\mr{\,erg\,s^{-1}} {\Md \over 3\times 10^{-3}\msun} {2.6\msun\over M} \lrb{\omega_{\rm BH}\over 0.4}^2 \\
    &\lrb{\PhiBH\over 0.3 \Phid}^2 \lrb{\theta\over 0.5}^3 \lrb{\xi P_{\rm B}\over 0.01P},
\end{split}
\end{equation}
where we have adopted some fiducial values of the ratios $\PhiBH/\Phid=0.3$ and $\xi P_{\rm B}/P=0.01$. These values have large uncertainties meaning that our eq. (\ref{eq:LBZ}) does not have strong predictive power and may only serve as a consistency check. An alternative model for the magnetic flux threading the horizon is provided by \citet{kisaka15_fallback_power}, who proposed that the BH's field topology evolves due to interactions with the fallback gas. A jet power of $L_{\rm BZ}\lesssim 10^{49}\rm\, erg\, s^{-1}$ lasting for $10^2\rm\, s$ would give a kinetic energy of $E_{\rm K}\lesssim 10^{51}\rm\, erg$, which is consistent the observational constraints from the afterglow\footnote{We note that it is often difficult to obtain the beaming-corrected GRB jet kinetic energy from the sparse afterglow observations in most cases \citep{berger14_sGRB_review, kumarzhang15_GRB_review} \citep[but see][]{beniamini15_prompt_efficiency}: one needs to measure the synchrotron cooling frequency to break the degeneracy between $\epsilon_{\rm B}$ (fraction of internal energy in the shock-heated region shared by B-fields) and the pre-shock gas density, and one also needs to detect the ``jet break'' in the late-time (faint) phase of the afterglow to constrain the opening angle of the jet. } and the fluence ratio between the EE and the prompt $\gamma$-ray emission (see eq. \ref{eq:fluence_ratio}). 


On the other hand, the jet power is limited by the maximum value expected from the limit of a Magnetically-Arrested Disk (MAD) where the magnetic pressure near the horizon is comparable to the ram pressure of the inflow \citep{narayan03_MAD, tchekhovskoy11_MAD_efficiencies, narayan22_MAD_efficiencies}, and this gives
\begin{equation}\label{eq:LMAD}
    L_{\rm MAD}\sim \dotMBH c^2\simeq 1.8\times10^{49}\mr{\,erg\,s^{-1}} {\dotMBH\over 10^{-5}\msun\rm\,s^{-1}},
\end{equation}
where $\dotMBH$ is the mass accretion rate onto the BH. However, due to possible wind mass loss, we expect the accretion rate at the horizon $\dotMBH$ to be a fraction $\eta_{\rm acc}<1$ of the mass inflow rate at the outer disk, i.e.,
\begin{equation}
    \dotMBH = \eta_{\rm acc}\Md/\tvis.
\end{equation}
The accretion fraction $\eta_{\rm acc}$ is currently highly uncertain due to the lack of global simulations that reach inflow equilibrium in a sufficiently large radial dynamical range, especially in the case where the BH launches a strong jet.
A physical testbed is provided by tidal disruption events (TDEs) with super-Eddington accretion and relativistic jets \citep{bloom11_jetted_TDE, cenko12_jetted_TDE}, where the observed X-ray lightcurve roughly follows the theoretically expected mass fallback rate of $\dot{M}_{\rm fb}\propto t^{-5/3}$. In these jetted TDE cases, the total jet energy of the order $10^{53}\rm\, erg$ \citep[as inferred from the radio afterglow,][]{decolle_jetted_TDE} indicates that a large fraction $\eta_{\rm acc}\gtrsim 0.1$ of the stellar mass is indeed accreted by the BH.

For the fiducial parameters in eq. (\ref{eq:LBZ}), the jet power is sub-MAD with $L_{\rm BZ}\lesssim L_{\rm MAD}$ for an accretion fraction of $\eta_{\rm acc}\gtrsim 0.1$ \citep[in contrast to the MAD case proposed by][]{barkov11_MAD_jet_EE}. The shallow flux decay in the EE and X-ray plateau is consistent with the case that the BH always has sub-MAD magnetic flux such that $L_{\rm BZ}< L_{\rm MAD}$ --- in this case the jet power would be proportional to the disk mass $L_{\rm BZ}\propto \Md\propto t^{-1/3}$ (eq. \ref{eq:LBZ}) instead of the accretion rate (which would otherwise give $L_{\rm BZ}\propto t^{-4/3}$). If the large-scale poloidal B-fields in the outer disk are only gradually amplified on a long timescale of $\gtrsim 10\rm\, s$, then the decay of the jet power may be even shallower than $t^{-1/3}$.

\section{Late-time Kilonova Emission}\label{sec:kilonova}
In this section, we discuss the imprint on the late-time kilonova emission of a long-lived disk of mass $M_{\rm d,0}\in (10^{-3}, 10^{-2})\msun$ after nuclear recombination. The key prediction is that the evaporation of the disk at time $t_{\rm evap}\sim 10^2\rm\, s$ (due to radioactive heating) generates a tail of very slow ejecta that produces narrow line features in the kilonova spectrum --- these are potentially detectable by JWST and future Extremely Large Telescopes.




In the self-similar solution of eq. (\ref{eq:self_similar}), the late-time outer disk radius is given by $\rd \simeq (3t\alpha\theta^2\sqrt{GM})^{2/3}$ and hence the typical asymptotic velocity of the disk wind is
\begin{equation}
\begin{split}
    v(t) &\simeq \sqrt{GM\over \rd} \simeq \lrb{GM\over 3t\alpha \theta^2}^{1/3}\\
    &\simeq 0.01c\, \lrb{t\over 100\mr{\,s}}^{-1/3} \lrsb{{M\over 2.6\msun} {0.1\over \alpha} \lrb{0.5\over \theta}^2}^{1/ 3}.
\end{split}
\end{equation}
Since the disk mass evolves as $\Md\propto t^{-1/3}$, we expect a slow ejecta tail of mass $\Mej \simeq (\tevap/t_0)^{-1/3}M_{\rm d,0}\simeq M_{\rm d,0}/3$ and velocity $v \simeq 0.01 c$ (for typical values $\tevap\sim 100\rm\, s$ and $t_0\sim 5\rm\, s$) as a result of late-time disk evaporation. In the following, we discuss the late-time kilonova signatures from this slow ejecta, which has not been considered in the literature before. 

We consider a representative model for the late-time disk evolution with $M_{\rm d,0}=3\times10^{-3}\msun$, $t_0= 5\rm\, s$, $\alpha=0.1$ and including wind cooling (this case corresponds to the green dotted line in Fig. \ref{fig:disk_evolution}). Starting from the time evolution of disk mass $\Md(t)$, radius $\rd(t)$ and viscous time $\tvis(t)$, we assume that a constant fraction $\eta_{\rm acc}=0.5$ of the accretion rate $\Md/\tvis$ is consumed by the BH and that the rest is ejected as wind at a mean velocity $v(t) = \sqrt{GM/\rd(t)}$. Our results are insensitive to the precise value of $\eta_{\rm acc}$.

\begin{figure}
  \centering
\includegraphics[width = 0.48\textwidth]{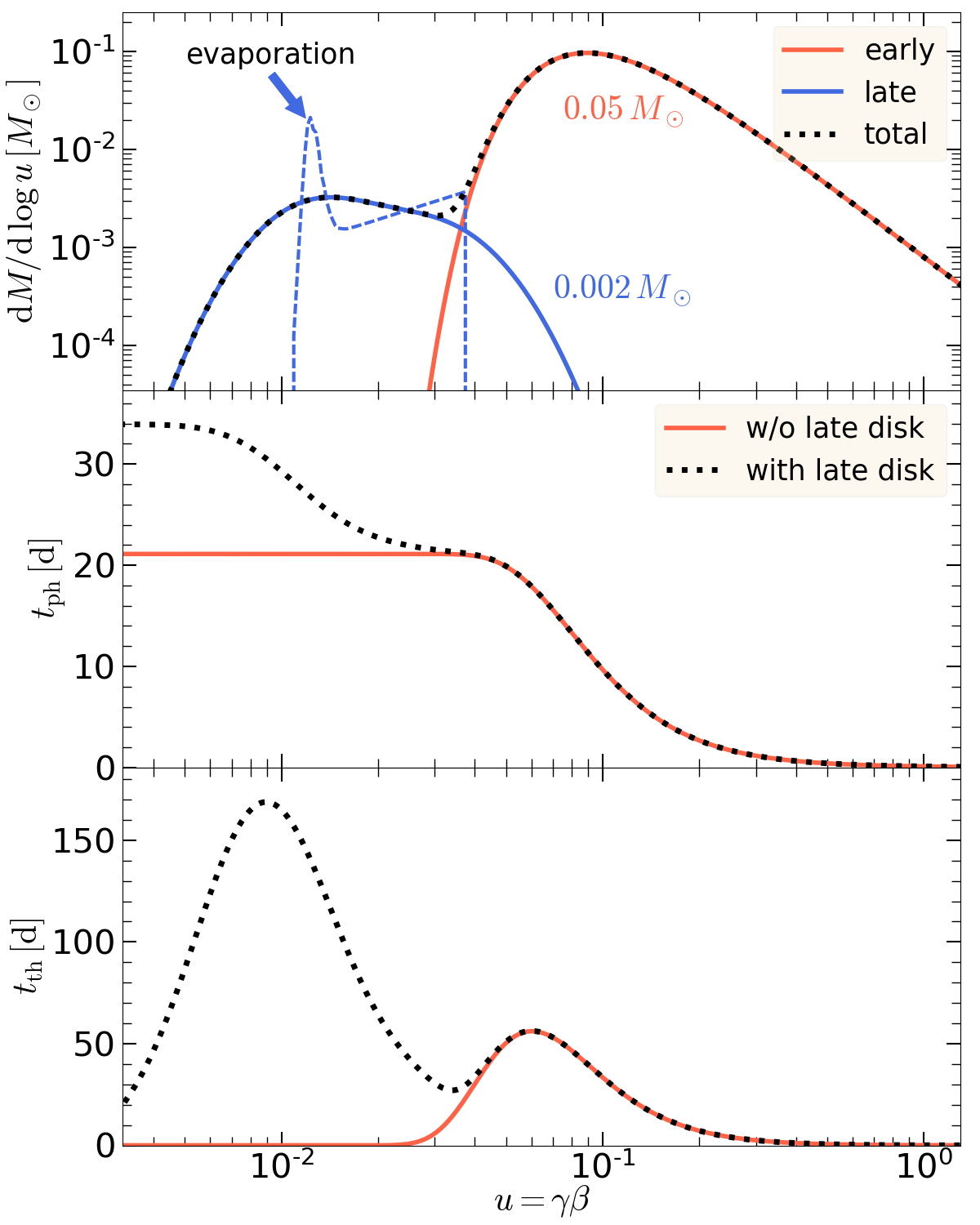}
\caption{\textit{Upper panel}: the ejecta velocity distribution, including the early/fast ejecta (red solid line) and late/slow ejecta (blue dashed/solid lines). The horizontal axis is the ejecta four-velocity in units of $c$. The early ejecta has a power-law four-velocity profile $\d M/\d \log u\propto u^{-2.5}$ at $u\gtrsim u_{\rm min}=0.1$ with a total mass of $0.05\msun$. The late ejecta's velocity distribution is obtained based on the disk evolution in the greed dotted line in Fig. \ref{fig:disk_evolution} (with $M_{\rm d,0}=3\times10^{-3}\msun$, $\alpha=0.1$, $t_0=5\rm\,s$), and we further assume that the BH's accretion rate is given by $\dotMBH(t)=0.5\Md/\tvis$ and that the rest of the disk mass ($\approx 2\times10^{-3}\msun$, including the radioactively driven evaporation) is carried away by the wind at a mean velocity $v(t)= \sqrt{GM/\rd(t)}$. The blue dashed line is for the case that the wind has a single speed at the mean value at a given time, and the blue solid line includes a broader instantaneous velocity distribution by a factor of $\sqrt{2}$ above and below the mean. \textit{Middle panel}: the time $t_{\rm ph}$ at which the photosphere recedes to a certain velocity layer. \textit{Bottom panel}: the time $t_{\rm th}$ at which $\beta$-decay electrons thermally decouple from the ejecta and their heating becomes less efficient. The slow ejecta stays optically thick for longer time (up to $34\rm\, d$) and are efficiently heated by $\beta$-electrons for a longer time (up to $190\rm\,d$) than the fast ejecta.
}\label{fig:dMdlgv}
\end{figure}

This allows us to calculate the mean velocity distribution of the late-time disk wind
\begin{equation}
    {\d M_{\rm late} \over \d v} = \lrb{\abs{\d \Md\over \d t} - \eta_{\rm acc}{\Md\over \tvis} } \abs{\d t \over \d v},
\end{equation}
which is shown by the blue dashed line in the upper panel of Fig. \ref{fig:dMdlgv}. In reality, the disk wind at a given time likely has a broader velocity distribution above and below the mean value, so we use a Gaussian filter to smooth out the mean velocity distribution by a standard deviation of $\sigma_{\log v}=0.15$ (corresponding to a factor of $\sqrt{2}$ in both directions), and the result is shown by the blue solid line in the upper panel of Fig. \ref{fig:dMdlgv}. Our qualitative results do not depend on the detailed smoothing procedure, since the mass in the slow ejecta is given by the disk evolution.

The peak kilonova emission is dominated by the much faster ($v\gtrsim 0.1c$) ejecta that is produced at earlier time ($t\lesssim\,$a few seconds) by tidal disruption and disk outflow driven by nuclear recombination. We schematically represent this faster ejecta component by the following power-law velocity distribution \citep[following][]{metzger19_kilonova_review}
\begin{equation}\label{eq:dMdlgv}
    u{\d M_{\rm early}\over \d u} = {2M_{\rm ej,early}\over \Gamma(p/2)} \lrb{u\over u_{\rm min}}^{-p} \mr{e}^{-u_{\rm min}^2/u^2},
\end{equation}
where $u=\gamma\beta$ is the four-velocity ($\beta=v/c$ and $\gamma$ is the Lorentz factor), $M_{\rm ej,early}=\int \d u (\d M_{\rm early}/\d u)$ is the total baryonic mass of the early ejecta, $u_{\rm min}$ is the minimum four-velocity below which the distribution cuts off, $p$ is the power-law index describing the steepness of the velocity distribution, and $\Gamma(x)$ is the Gamma function. We take $u_{\rm min}=0.1$ and $p=2.5$ such that most of the kinetic energy is contained in the slower parts near $u_{\rm min}$ as motivated by the disk simulations by \citet{siegel18_GRMHD_disk, fernandez19_long_GRMHD}. The highly relativistic ejecta with $u\gtrsim 1$ as well as the ultra-relativistic jets are unimportant for our purpose here since we focus the late-time kilonova emission. The early ejecta mass is taken to be $M_{\rm ej,early}=0.05\msun$, as motivated by kilonova lightcurve of GW170817 \citep[see][for a number of best-fit models]{wu19_late_time_lightcurve}. The resulting velocity distribution of the early ejecta is shown by the solid red line in the upper panel of Fig. \ref{fig:dMdlgv}.

With the velocity distribution, we consider the whole ejecta to be in homologous expansion like the Hubble flow. The time $t_{\rm ph}(u)$ at which the photosphere recedes to a certain velocity layer is obtained by solving
\begin{equation}
    \int_u^\infty \d u {\d M\over \d u} {\kappa \over 4\pi \beta^2 c^2 t_{\rm ph}^2} = 1,
\end{equation}
where $\kappa$ is the Planck-mean opacity. The realistic opacity depends on chemical composition, electron temperature, and ionization states of each species \citep{barnes13_high_opacity, tanaka13_high_opacity, kasen17_kilonova_modeling, fontes20_opacity, tanaka20_opacity, hotokezaka21_nebular_emission, pognan22_nebular_opacity}. We take a constant opacity of $\kappa=3\rm\, cm^2\,g^{-1}$, as representative for intermediate electron fraction $\Ye\in(0.25, 0.35)$ \citep[see][]{tanaka20_opacity}. For the longer expansion times of $\mc{O}(1)\rm\,s$ appropriate to the disk viscous evolution, the mass fractions of Lanthanides and Actinides at a given $\Ye$ are somewhat lower than for the shorter expansion times typically considered for NS merger dynamical ejecta or disk winds (See Fig. \ref{fig:yfinal}). A proper calculation of the late-time slow ejecta's opacity will require modeling of the disk's expansion history coupled to r-process nucleosynthesis.

The results of $t_{\rm ph}(u)$ are shown in the middle panel of Fig. \ref{fig:dMdlgv}. We find that the late-time slow ejecta from disk evaporation extends the photospheric phase by about 10 days as compared to the case without a long-lived disk.

How much does the slow ejecta contribute to the late-time continuum and line emission? We provide some crude estimates in the following.


Let us assume that the late-time heating is entirely due to $\beta$-decays, although $\alpha$-decays and fission might also contribute \citep[provided that the nucleosynthesis produces nuclei beyond the third r-process peak, see][]{wu19_late_time_lightcurve}.
After injection, a $\beta$-electron loses energy due to adiabatic expansion $\dot{E}_{\rm e,ad} = -(1+1/\ge)\Ee/t$ and collisional losses $\dot{E}_{\rm e,coll}=-K_{\rm st}\be \rho c$, and these two terms can be manipulated such that its four-velocity $\ue\equiv \ge \be$ evolves according to
\begin{equation}\label{eq:electron_cooling}
    {\d \ue \over \d t} = -{\ue \over t} - {K_{\rm st}\rho \over \me c},
\end{equation}
where $t$ is the time since explosion, $\be=v_{\rm e}/c$ is the dimensionless speed ($\ge=1/\sqrt{1-\be^2}$ being the Lorentz factor), $\rho$ is the local ejecta density, $\me$ is electron rest mass, and $K_{\rm st}$ is the stopping coefficient given by
\begin{equation}
    K_{\rm st} \simeq (1.2 \ue^{-1.3} + 0.5\ue^{0.5})\, \mr{MeV\,cm^2\, g^{-1}}.
\end{equation}
The above stopping coefficient is a numerical fit based on analytical expressions including ionization losses, Coulomb interactions with other electrons, and Bremstruhlung by heavy nuclei \citep[e.g.,][]{barnes16_heating_thermalization, waxman19_thermalization, hotokezaka20_thermalization}.

There is a critical time $t_{\rm th}$ at which the $\beta$-electrons thermally decouple with the local ejecta, and it is given by
\begin{equation}
    {\ue\over t_{\rm th}}\simeq {K_{\rm st}\rho(t_{\rm th}) \over \me c},
\end{equation}
where the ejecta density of a layer at bulk four-velocity $u$ is given by
\begin{equation}
    \rho(u, t) = {\d M/\d v\over 4\pi v^2 t^3} = {\d M/\d u \over \gamma u^2 4\pi c^3 t^3}.
\end{equation}
Thus, the $\beta$-electron decoupling time is given by
\begin{equation}
    t_{\rm th}(u) \simeq \lrb{{K_{\rm st}\over \ue \me c^2 4\pi c^2} {\d M /\d u \over \gamma u^2}}^{1/2},
\end{equation}
where we take $\ue \simeq 2$ and $K_{\rm st}\simeq 1.2\mr{\,MeV\, cm^2\,g^{-1}}$ (corresponding to a typical injection kinetic energy of $\Ee\simeq 0.5\,\rm MeV$).

At sufficiently late time $t\gtrsim 10\rm\, d$ when photon trapping is negligible, the kilonova bolometric luminosity (ignoring the X-ray and $\gamma$-ray photons) is equal to the sum of the heating rates $q_{\rm e}(u, t)$ contributed by all fluid elements, i.e.,
\begin{equation}
    L_{\rm bol}(t) = \int \d u {\d M\over \d u} q_{\rm e}(u, t),
\end{equation}
Following \citet{hotokezaka20_thermalization}, we take the solar r-process abundance pattern as the final composition of stable elements with minimum and maximum atomic mass numbers $A_{\rm min}=85$ and $A_{\rm max}=209$, and hence the heating rate from $\beta$-electrons can be approximated by
\begin{equation}\label{eq:qe}
    q_{\rm e}(u, t) \simeq 4\times10^9\mr{erg\over s\,g} \lrb{t/\mr{day}}^{-1.3} \lrb{1 + {t\over t_{\rm th}(u)}}^{-1.5},
\end{equation}
where the asymptotic behavior of $q_{\rm e}\propto t^{-2.8}$ after thermal decoupling is due to the following reason \citep[see][]{waxman19_thermalization}. At late time $t\gg t_{\rm th}$,
the heating rate $q_{\rm e}=\int \d \ue \lrb{\d N/\d \ue} K_{\rm st}\be \rho c$ is dominated by the electrons near velocity $\ue\sim t_{\rm th}/t$ (these were injected at time $t\sim t_{\rm th}$), and the ejecta expansion $\rho\propto t^{-3}$ plus the weak dependence of $K_{\rm st}\be$ on electron velocity give rise to a roughly power-law behavior $q_{\rm e}\propto t^{-2.8}$.

\begin{figure}
  \centering
\includegraphics[width = 0.48\textwidth]{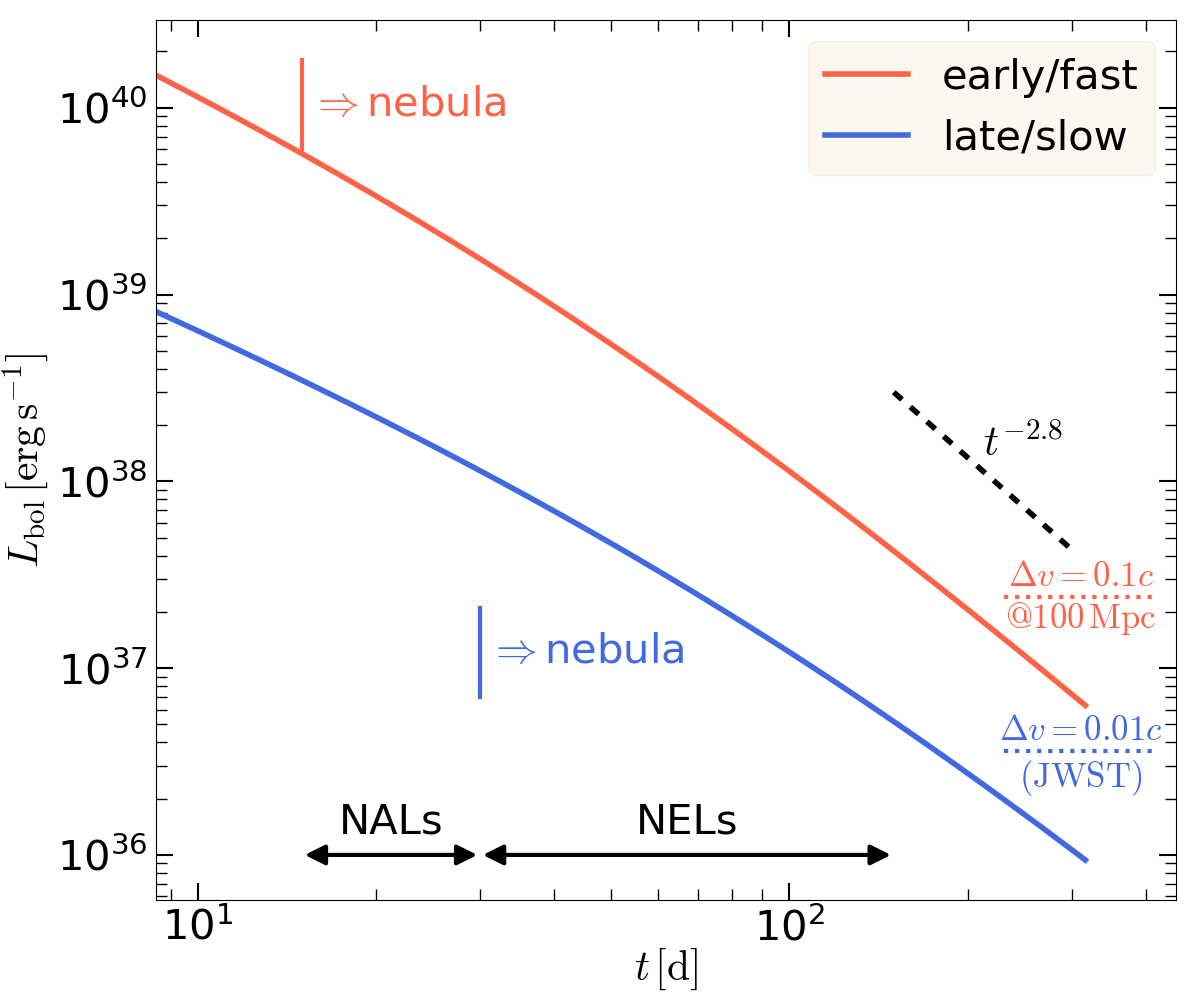}
\caption{The late-time bolometric lightcurve for the ejecta velocity distribution in Fig. \ref{fig:dMdlgv}, including contributions from both the early/fast (red) and late/slow (blue) ejecta components. Only $\beta$-decay electron heating is included, at a rate (eq. \ref{eq:qe}) given by an ejecta composition that will reproduce the solar r-process abundance pattern via $\beta$-decays. The early ejecta produces a photospheric continuum with broad ($\Delta v\sim 0.1 c$) absorption lines at $t\lesssim 15\rm\, d$ and then broad emission lines at $t\gtrsim 15\rm\, d$. The late ejecta produces a photospheric continuum with narrow absorption lines (NALs) of widths $\Delta v\sim 0.01 c$ at $t\lesssim 30\rm\, d$ and then narrow emission lines (NELs) later. The higher density late ejecta has longer-lasting efficient thermalization than the early ejecta, which increases the fractional luminosity contribution by the late ejecta to $\sim\, 10\%$ at $t\gtrsim 100\rm\, d$. The horizontal dotted lines show  line-luminosity sensitivities ($\rm SNR=5$ at line center) for 10 ks JWST exposure for broad (red) and narrow (blue) lines and for a source distance of $100\rm\, Mpc$.
}\label{fig:Lbol}
\end{figure}

The contributions to the bolometric luminosity by the early/fast and late/slow ejecta components are shown in Fig. \ref{fig:Lbol}. We find that the slow ejecta from the late-time evaporation of the disk contributes $\sim\!10\%$ of the bolometric luminosity at very late time $t\gtrsim 100\rm\, d$. Such a small contribution is difficult to identify photometrically. The \textit{Spitzer} mid-infrared (mean wavelength $\lambda\approx 4.5\rm\,\mu m$) detections of the GW170817 kilonova at $t=43$ and $74\rm\, d$ \citep{kasliwal22_spitzer_GW170817} is almost certainly driven by the dominant fast ejecta since the IRAC Channel 2 photometric band is wide $\Delta \lambda \sim 1\rm\,\mu m$. However, the contribution from the slow $v\sim 0.01 c$ ejecta could be disentangled via spectroscopy.





For a 10 ks exposure time, JWST low-resolution ($\lambda/\Delta\lambda \sim 100$) fixed-slit spectroscopy using NIRSpec with PRISM disperser and CLEAR filter (0.6--5.3$\rm \mu m$) has $5\sigma$ line flux sensitivity\footnote{This is for $\rm SNR=5$ at the line center, based on JWST Exposure Time Calculator at \href{https://jwst.etc.stsci.edu}{https://jwst.etc.stsci.edu}} in the wavelength range of $3\times10^{-18}\rm\, erg\, s^{-1}\, cm^{-2}$ for a narrow line of width $\Delta v = 0.01c$ and $2\times10^{-17}\rm\, erg\, s^{-1}\, cm^{-2}$ for a broad line of width $\Delta v = 0.1c$. These line flux limits are shown as horizontal dotted lines in Fig. \ref{fig:Lbol} for a source at a distance of 100 Mpc.

At time $15\lesssim t\lesssim 30\rm\, d$, the faster ejecta is optically thin, and this allows the observer to see the emission from the slower ejecta in the deep interior. It is possible to identify narrow absorption lines (NALs) at wavelengths where the flux contribution from the broad emission lines from the faster ejecta is minimized (so as to reduce the dilution for the NALs). In addition to JWST, the slower ejecta's photospheric continuum at $\lambda\lesssim 2\rm\, \mu m$ will also be accessible to ground-based spectroscopic observations by future Extremely Large Telescopes. At later time $t\gtrsim 30\rm\, d$, the slower ejecta becomes optically thin as well, and since the emission shifts to longer wavelengths, space-based observations are required. The late-time spectrum is expected to show narrow emission lines (NELs) whose strengths are not affected by broad emission lines from the faster ejecta.  Identification of atomic lines will provide powerful diagnostic for the chemical composition of the ejecta \citep[see][for potential identification of Strontium in the early kilonova spectra]{watson19_strontium_line}. 

Finally, we note that, even without the late-time disk evaporation, the early ejecta (launched at $t\lesssim\,$a few seconds) may already contain a slow tail at $v\ll 0.1c$ as a result of energy exchange between fluid elements due to pressure gradient, which may produce a power-law instead of an exponential cut-off at the low-velocity end of the velocity distribution in eq. (\ref{eq:dMdlgv}). In this case, the total mass of low-velocity ejecta near $v\sim 0.01c$ would be even higher than considered in this section and the predicted late-time narrow line features would be more prominent.

\section{Discussion --- Constraints on Two-component Jet}\label{sec:discussion}

Our model predicts that the relativistic jet from a NS merger has two components: one is launched in the first $\mc{O}(1)\rm\, s$ and produces the prompt $\gamma$-ray emission; the other one is produced by long-lived BH accretion which lasts for $\mc{O}(10^2)\rm\, s$ and is responsible for the EE/X-ray plateau. These two components should in principle have different opening angles $\Omega_{\gamma}$ and $\Omega_{\rm EE}$, although it is currently difficult to predict these two values from our model and hence they can only be constrained by observations --- observers at different viewing angles would see a diverse set of phenomena in the $\gamma$-ray or hard X-ray band at early time (when the jet undergoes internal dissipation) and in the broad-band afterglow emission at late time (when the jet interacts with the interstellar medium).

In this section, we discuss the constraints on our model by the observed fluence (or isotropic-equivalent energy) ratio between the EE and the prompt $\gamma$-ray emission in the \swift BAT band (15--150 keV), which spans a wide range from $\mc{F}_{\rm EE}/\mc{F}_{\gamma}\lesssim 0.1$ to $\sim\!30$ \citep[see Fig. 4 of][]{perley09_sGRB_EE}. This is perhaps not surprising, as there are a number of highly uncertain factors that could cause the variations of this ratio from burst to burst.

Assuming that our line of sight is inside the beaming cones of both EE and the prompt $\gamma$-ray emission, we write
\begin{equation}\label{eq:fluence_ratio}
    {\mc{F}_{\rm EE}\over \mc{F}_{\gamma}} = {\epsilon_{\rm EE}\over \epsilon_{\gamma}} {\Omega_{\gamma}\over \Omega_{\rm EE}} {L_{\rm j, EE}\over L_{\rm j,\gamma}} {T_{\rm EE} \over T_\gamma},
\end{equation}
where $L_{\mr{j},i}$, $T_{i}$, $\Omega_i$ and $\epsilon_i$ are the jet luminosity, duration, beaming solid angle, and radiative efficiency in the \swift BAT band for the two components ($i=\mr{EE}$ or $\gamma$). Apart from $T_{\rm EE}/T_\gamma\sim 10^2$, we briefly discuss the other factors in the following.

Compared to the prompt emission, the EE usually has a softer spectrum such that a larger fraction of the energy may be emitted below $\sim\! 20\rm\,keV$ where \swift BAT loses effective area significantly, so we might expect $\epsilon_{\rm EE}/\epsilon_\gamma < 1$. By modeling the afterglow emission, one can constrain the total isotropic equivalent kinetic energy $E_{\rm K,iso}$, provided that the magnetization of the shocked emitting region is known (by measuring the synchrotron cooling frequency). Under the \textit{assumption} that the jet responsible for the prompt emission dominates $E_{\rm K,iso}$, previous works obtained $\epsilon_\gamma\sim 0.1$ for some  GRBs with bright afterglows \citep{beniamini15_prompt_efficiency, beniamini16_prompt_efficiency}, but since the relative fractional contributions from EE and prompt emission to $E_{\rm K,iso}$ is currently unknown, this exercise does not strongly constrain the ratio of $\epsilon_{\rm EE}/\epsilon_\gamma$ \citep[but see][for a model-dependent constraint]{matsumoto20_EE_efficiency}.



Observational constraints on $\Omega_{\gamma}$ can be obtained as follows. Under the (perhaps reasonable) assumption that each NS merger generates a short GRB, one can compare the NS merger rate of $10^2$--$10^3\rm\, Gpc^{-3}\, yr^{-1}$ \citep{abbott21_LIGO_merger_rate} to the on-axis short GRB rate $2$--$6\rm\, Gpc^{-3}\,yr^{-1}$ \citep[obtained by modeling their luminosity function, e.g.,][]{wanderman15_sGRB_luminosity_function} --- this roughly gives $\Omega_\gamma/4\pi \sim 10^{-2}$ but with large uncertainties. On the other hand, one can measure the jet opening angle $\theta_{\rm j}$ by searching for the ``jet break'' signature \citep{sari99_jet_lateral_expansion} in the multiband afterglow lightcurves. However, despite several likely detections, many short GRBs do not show a jet break before the afterglow fades below the detection threshold. \citet{berger14_sGRB_review} reviewed these observations and concluded that the mean opening angle $\lara{\theta_{\rm j}}\gtrsim 10^{\rm o}$ corresponding to $\Omega_{\rm j}/4\pi \gtrsim 1.5\times10^{-2}$. Here, the total jet solid angle may be understood as $\Omega_{\rm j} = \max(\Omega_\gamma, \Omega_{\rm EE})$ (although this depends on the isotropic-equivalent energies of the two components). We conclude that the case of $\Omega_{\rm EE} > \Omega_\gamma$ is allowed by current observations\footnote{\citet{barkov11_MAD_jet_EE} proposed a two-component jet model with $\Omega_\gamma > \Omega_{\rm EE}$ (their Fig. 3), where the prompt $\gamma$-ray jet component is produced by the neutrino annihilation mechanism (with a wider beaming angle) and the EE jet component is produced by the Blandford-Znajek process (with a narrower beaming angle). In this paper, we present arguments suggesting the opposite --- $\Omega_{\rm EE}>\Omega_\gamma$. This can be tested by the data from future $\gamma$/X-ray surveys on the rate of short GRBs and EEs (including ``orphan EEs''). } and even favored by some short GRBs lacking a jet break up to several weeks after the burst \citep[e.g.,][]{grupe06_no_jet_break, berger13_no_jet_break}. Recently, \citet{laskar22_late_jet_break} reported the detection of a very late jet break from the short GRB 211106A at observer's time $\sim\!30\rm\, d$, which implies $\theta_{\rm j}\sim 15^{\rm o}$ or $\Omega_{\rm j}/4\pi\sim 0.03$.

The afterglow emission from a two-component jet model has been considered by \citet{huang04_2component_jet, peng05_2component_jet} who show that such a model explains the late-time afterglow data in many GRBs better than a single-component jet.
The picture consists of two jet components with (true) kinetic energies $E_1$ and $E_2$ and opening angles $\theta_1 <\theta_2$ --- the second component usually has a lower isotropic equivalent energy $E_2/\theta_2^2 < E_1/\theta_1^2$. For an observer on the jet axis with viewing angle $\theta_{\rm obs}<\theta_1$, the wider jet component with lower isotropic equivalent energy can dominate the afterglow emission after the jet-break time of the narrower component \citep{peng05_2component_jet}. For an observer far from the jet axis $\theta_{\rm obs}>\theta_2$, as is the case for GW170817 \citep{mooley18_VLBI_proper_motion}, the wider jet component decelerates faster and its emission comes into our view earlier than the narrower jet component. The afterglow emission from component $i=1, 2$ (at a given observer's frequency of e.g., 10 GHz) reaches a peak flux of $F_{\mr{p}, i}\propto E_i^{(p+1)/4}$ at time $t_{\mr{p}, i}\propto E_i^{1/3} (\theta_{\rm obs}-\theta_i)^2$ \citep{nakar02_offaxis_afterglow, gottlieb19_offaxis_afterglow}. We see that the emission from the wider jet component likely peaks only slightly earlier than that from the narrower jet --- $t_{\rm p,2}/t_{\rm p,1}$ is of order unity due to the weak dependence on $E_i$. However, the peak flux $F_{\rm p}$ has a strong dependence on the jet energy --- and we find that the afterglow data of GW170817 \citep{margutti21_GW170817_ARAA} cannot rule out a two-component jet with $E_2/E_1\lesssim 1/3$.


The intriguing possibility of  $\Omega_{\rm EE} > \Omega_\gamma$ allows for the detection of ``orphan EE'' without a prompt $\gamma$-ray spike (in this case $\mc{F}_{\rm EE}/\mc{F}_{\gamma}\rightarrow \infty$). Some supernova-less apparently long-duration GRBs \citep{gehrels06_GRB060614, rastinejad22_GRB211211A, yang22_GRB211211A} may indeed be such cases, although the rate of these events are highly uncertain. Other possible cases are the so-called X-ray Flashes (XRFs), which have much softer spectra than GRBs \citep{sakamoto05_XRF}. Furthermore, the population of ultra-long GRBs with durations of $10^3$--$10^4\rm\,s$ \citep{gendre13_ULGRB_111209A, levan14_ULGRB, greiner15_ULGRB_111209A} may be the ``orphan EE'' from normal long GRBs from standard collapsars (without requiring an extended blue supergiant).

Since \swift BAT is less likely to trigger on faint-long events than on bright-short ones and is not sensitive below $\sim\!20\rm\, keV$, it does not provide a very strong constraint on $\Omega_{\rm EE}$ by itself. Nevertheless, the case of $\Omega_{\rm EE}/4\pi \gtrsim 0.1$ can be ruled out by current non-detections, because that would produce $\gtrsim 1$ event within 400 Mpc every year in the $1.4\rm\,sr$ field of view of \swift BAT (for a NS merger rate of $300 \rm\, Gpc^{-3} \,yr^{-1}$). For a modest isotropic equivalent energy of $10^{50}\rm\, erg$, each of these events would produce a fluence of $5\times 10^{-6}\rm\, erg\,cm^{-2}$, which is easily detectable by BAT \citep[see Fig. 11 of][]{levan14_ULGRB}.

A physical reason that may lead to  $\Omega_{\rm EE} > \Omega_\gamma$ is as follows. In NS-NS mergers, there is a significant amount of baryonic material (e.g., the dynamical ejecta) near the rotational axis of the system. Thus, the prompt $\gamma$-ray jet must push its way through the surrounding gas --- a high-pressure cocoon surrounding the jet is produced in this process and the cocoon provides additional collimation for the early jet \citep{ramirez-ruiz02_cocoon, bromberg11_cocoon, nagakura14_jet_collimation, duffel18_jet_propagation, geng19_HDjet_propagation, gottlieb21_HDjet_cocoon, gottlieb22_MHDjet_cocoon}. In the case of a successful jet break-out, the dynamical ejecta gas near rotational axis is evacuated as a result of lateral expansion of the shocked gas. Then, the late EE jet is likely less collimated due to a weaker cocoon (though the late-time disk wind can still provide some collimation). Detailed hydrodynamic simulations including a long-lived jet component are needed to provide a better prediction on $\Omega_\gamma$ and $\Omega_{\rm EE}$.  In this scenario, it is likely that NS-BH mergers will have more similar $\Omega_\gamma$ and $\Omega_{\rm EE}$ because they lack the polar ejecta that provides most of confining pressure in NS-NS mergers.

Finally, the jet luminosity ratio $L_{\rm j, EE}/L_{\rm j, \gamma}$ likely differs substantially from source to source due to the variation in disk mass $M_{\rm d,0}$ at the end of nuclear recombination. The global GRMHD simulations by \cite{siegel18_GRMHD_disk, fernandez19_long_GRMHD} show that of the order 1\%--10\% of the disk mass right after merger stays bound at the end of helium recombination. This suggests that the disk mass ratio alone would give $L_{\rm j, EE}/L_{\rm j, \gamma}\sim 0.01$ to $0.1$. The other factors in eq. (\ref{eq:LBZ}) $\PhiBH/\Phid$ and $\xi P_{\rm B}/P$ (which depends on poorly understood MHD processes) may also cause fluctuations in the jet luminosity ratio.

Based on the above considerations, we conclude that currently existing observations are consistent with the scenario that the observed EE is powered by a long-lived accretion disk with mass $M_{\rm d,0}\in (10^{-3},10^{-2})\msun$ that remains after nuclear recombination. The short GRBs without EE detections may be due to softer spectrum that is below the energy band of Swift BAT, a wide beaming angle $\Omega_{\rm EE}$ that dilutes the EE flux, very low leftover mass $M_{\rm d,0}$, inefficient accumulation of magnetic flux at the BH horizon, or a combination of these factors.

Fortunately, the ECLAIRs telescope onboard the SVOM mission is sensitive down to 4 keV \citep{godet14_ECLAIRS} and will likely detect a large sample of EEs either accompanied by a short GRB or without a prompt $\gamma$-ray spike (i.e. ``orphan EE'', provided that $\Omega_{\rm EE}>\Omega_\gamma$). Other wide-field X-ray surveys like Einstein Probe \citep{yuan15_Einstein_Probe} and THESEUS \citep{cielfi21_Theseus} will also detect many EEs. These instruments will provide more accurate measurements of the EE spectrum, a better understanding of the statistical distribution of $\mc{F}_{\rm EE}/\mc{F}_{\gamma}$, as well as a tighter constraint on $\Omega_{\rm EE}$ in the near future.



\section{Summary}\label{sec:summary}

This work is motivated by (i) the need to understand the long-term ($t\gg 10\rm\, s$) evolution of the accretion disk in NS mergers, (ii) the puzzling observations of many short GRBs showing extended emission (EE) and X-ray plateau lasting for $t\sim 10^2\rm\, s$ followed by a sharp drop in $\gamma/$X-ray flux, (iii) future JWST spectroscopic observations of kilonovae especially at late time ($t\gtrsim 15\rm\, d$).

We construct a model for the late-time disk evolution taking into account the radioactive heating by r-process nuclei formed earlier in the nuclear recombination phase (the first few seconds), under the assumption that a fraction (1\% to 10\%) of the initial disk mass is left bound at the end of nucleosynthesis. As the disk viscously spreads in radius, its binding energy $\propto t^{-2/3}$ drops faster than the cumulative heating by radioactive decays $\propto t^{-0.3}$ (provided that $\Ye\lesssim 0.4$ right before r-process nucleosynthesis). There is a critical time $\tevap$ when the disk material is overheated to become unbound and hence will quickly evaporate. We propose that the jet power rapidly drops and hence the EE and X-ray plateau ends abruptly at the evaporation time $\tevap$.

Our crude semi-analytic disk evolution model predicts $\tevap\sim 10^2\rm\, s (\alpha/0.1)^{-1.8} (M/2.6\,\msun)^{1.8}$ in the scenario without cooling by the disk wind (\S\ref{sec:without_wind}), and for the case with wind cooling (\S\ref{sec:with_wind}), an evaporation time of $10^2\rm\, s$ is obtained at a critical Bernoulli number of $B_{\rm e,0}=0.1$ (which determines the wind cooling rate by eq. \ref{eq:Be0}) and $\alpha=0.1$. Long GRBs from the collapse of massive stars likely produce more massive BHs and hence their X-ray plateaus should last longer ($\tevap\sim 10^4\rm\, s$ requires $M\sim 30\,\msun$).  However, the strong dependence of $\tevap$ on $\alpha$ and $B_{\rm e,0}$ means our model needs to be carefully calibrated against GRMHD simulations in order to be strongly predictive. 

We also propose a new scenario for the jet power that can potentially explain the shallow decay in the lightcurves of the EE and X-ray plateau. Recent simulations by \citet{liska20_disk_dynamo} showed that the $\alpha\omega$-dynamo operating in the outer disk produces poloidal B-fields that are coherent on the lengthscale of the vertical pressure scaleheight. We further demonstrate that if the magnetic pressure stays at a fixed fraction of the total pressure over the viscous evolution of the disk, then the total poloidal magnetic flux in the outer disk scales with the disk mass as $\Phid\propto \Md^{1/2}$. This motivates us to consider a jet launched by the \citet{blandford77_BZjet} mechanism (based on frame-dragging of the B-field lines threading the horizon) with a total power that scales as $L_{\rm BZ}\propto \Phid^2\propto \Md$. Before the radioactively driven evaporation, the disk mass evolves slowly with time $\Md\propto t^{-1/3}$, so this produces a slowly evolving jet power $L_{\rm BZ}\propto t^{-1/3}$ that is roughly consistent with observations, provided that the leftover disk mass at the end of the nuclear recombination is roughly in the range of $10^{-3}\lesssim M_{\rm d,0} \lesssim 10^{-2}\msun$. However, the leftover disk mass $M_{\rm d,0}$ at the end of r-process nucleosynthesis is currently rather uncertain. This uncertainty can be reduced by future simulations that dynamically couples MHD with nucleosynthesis beyond He. Making further predictions for the $\gamma$/X-ray lightcurves requires better understanding of the processes that determines the magnetic flux threading the BH horizon as well as the beaming angle and X-ray emission mechanism of jets.


Existing afterglow observations suggest that the beaming angle of the EE may be wider than that of the prompt $\gamma$-ray emission. We propose a possible physical reason for this: the earlier prompt jet is strongly collimated by the cocoon pressure whereas the EE jet propagates in the cavity produced by the expanding cocoon and is hence less collimated. This means that ``orphan EEs'' (without prompt $\gamma$-rays) may be promising electromagnetic counterparts to NS-NS or NS-BH mergers, and they are observable by future wide-field X-ray surveys, e.g., ECLAIRs \citep{godet14_ECLAIRS}, Einstein Probe \citep{yuan15_Einstein_Probe}, and THESEUS \citep{cielfi21_Theseus}. Possible existing examples of orphan EEs are the supernova-less apparently long-duration ($10^2\rm\,s$) GRBs whose ``prompt emission'' properties are similar to EE. Another potential support of our 2-component jet model is provided by the population of ultra-long GRBs whose ``prompt emission'' lasts for $10^4\rm\,s$ --- these could be orphan EE from normal long GRBs from collapsars.

Alternative scenarios for the EE/X-ray plateau are based on the spindown of a supra-massive NS or fallback accretion. These models, as well as our proposal here, all have large theoretical uncertainties. A major difference is that
the energy release in the EE phase is modest $\lesssim 10^{51}\rm\, erg$ in our model based on long-lived disk accretion whereas we expect a typical energy injection of $\gtrsim 10^{52}\rm\, erg$ in the supra-massive NS model. A larger energy injection produces much brighter afterglow emission when the kilonova ejecta is decelerated by the surrounding medium. Another prediction of our model is that short GRBs due to BH-NS mergers should also show EE/X-ray plateau followed by a rapid decline, and since the BH is likely more massive ($M\gtrsim 5\msun$), the plateau lasts for longer ($\gtrsim 300$ seconds) than in the NS-NS merger case. These predictions are testable by future multi-messenger observations of BH-NS and NS-NS mergers.

Finally, the late-time disk evaporation produces a tail of very slow ($\sim\!0.01c$) ejecta following the faster ($\gtrsim 0.1c$) ejecta generated by the tidal disruption and the early disk wind. We show that the late/slow ejecta, due to its higher density, can efficiently thermalize the energy injection from $\beta$-decay electrons (as well as the nuclear fragments from $\alpha$-decay and fission) up to $t\sim 100\rm\, d$, which increases its contribution to the bolometric luminosity to of the order $10\%$ at these late epochs. We predict that JWST NIRSpec observations of nearby ($\lesssim 100\rm\, Mpc$) NS mergers will be able to detect narrow ($\Delta v\sim 0.01c$) line features a few weeks after the merger. This will potentially provide a powerful probe of the detailed chemical composition of the r-process enriched ejecta (which is otherwise much harder to achieve if all lines are very broad and significantly overlap with each other).

\section*{Acknowledgments}
We thank Brian Metzger for suggesting looking into possible late-time line features in kilonovae. We thank Paz Beniamini for helpful discussion on the afterglow emission from two-component jets. We also thank Dan Kasen, Jim Stone, Todd Thompson, Bradley Meyer, Adam Burrows, Andrei Beloborodov, Matt Coleman, Bing Zhang, Pawan Kumar, Ryan Chornock, and Raf Margutti for useful discussions. WL was supported by the Lyman Spitzer, Jr. Fellowship at Princeton University. EQ was supported in part by a Simons Investigator grant from the Simons Foundation.

\section*{Data Availability}
The data underlying this article will be shared on reasonable request to the corresponding author.

{\small
\bibliographystyle{mnras}
\bibliography{refs}
}

\appendix
\section{Radioactive Heating}\label{sec:heating_rate}

Past numerical studies \citep[e.g.,][]{metzger09_Ye_freeze, siegel18_GRMHD_disk, fernandez19_long_GRMHD} showed that as the post-NS-merger disk evolves with time, the electron fraction of the disk material freezes at a moderately low value $\Ye\lesssim 0.4$. Observations of the kilonova from GW170817 suggest that the majority of the ejecta mass likely comes from a neutron-rich ($\Ye\lesssim 0.25$, as required by the long-lasting red kilonova emission component) accretion disk outflow \citep[][and refs therein]{metzger19_kilonova_review}.

\begin{figure}
  \centering
\includegraphics[width = 0.48\textwidth]{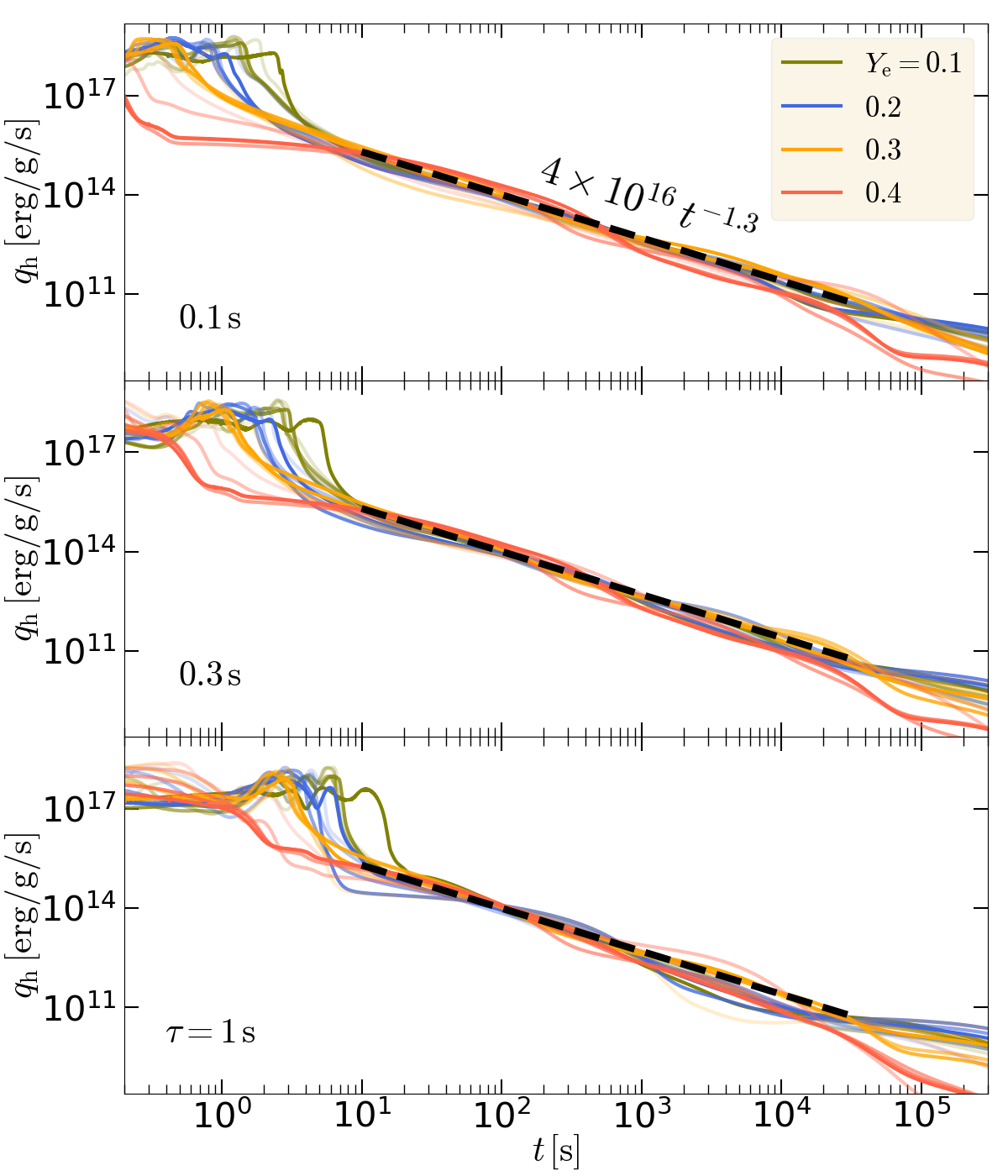}
\caption{Numerical heating rates for different expansion times $\tau=0.1\rm\, s$ (upper), $0.3\rm\,s$ (middle), $1\rm\, s$ (bottom panels). Different colors represent initial electron fractions of $\Ye=0.1$ (green), $0.2$ (blue), $0.3$ (orange), and $0.4$ (red), as expected from the viscous evolution of the accretion disk.  The initial conditions are taken to be NSE at temperature $T_0=6\rm\, GK$ and the initial density is set by the specific entropy $s_0$. For each $\Ye$, we consider 5 logarithmically spaced entropies between $s_0=8$ and $350\rm\, k_{\rm B}/\mp$, as shown in different shaded lines --- the lightest line corresponds to the highest specific entropy and hence lowest density. The black dashed lines in all three panels are identical as given by eq. (\ref{eq:qh}).
}\label{fig:heating_rates}
\end{figure}

\begin{figure}
  \centering
\includegraphics[width = 0.48\textwidth]{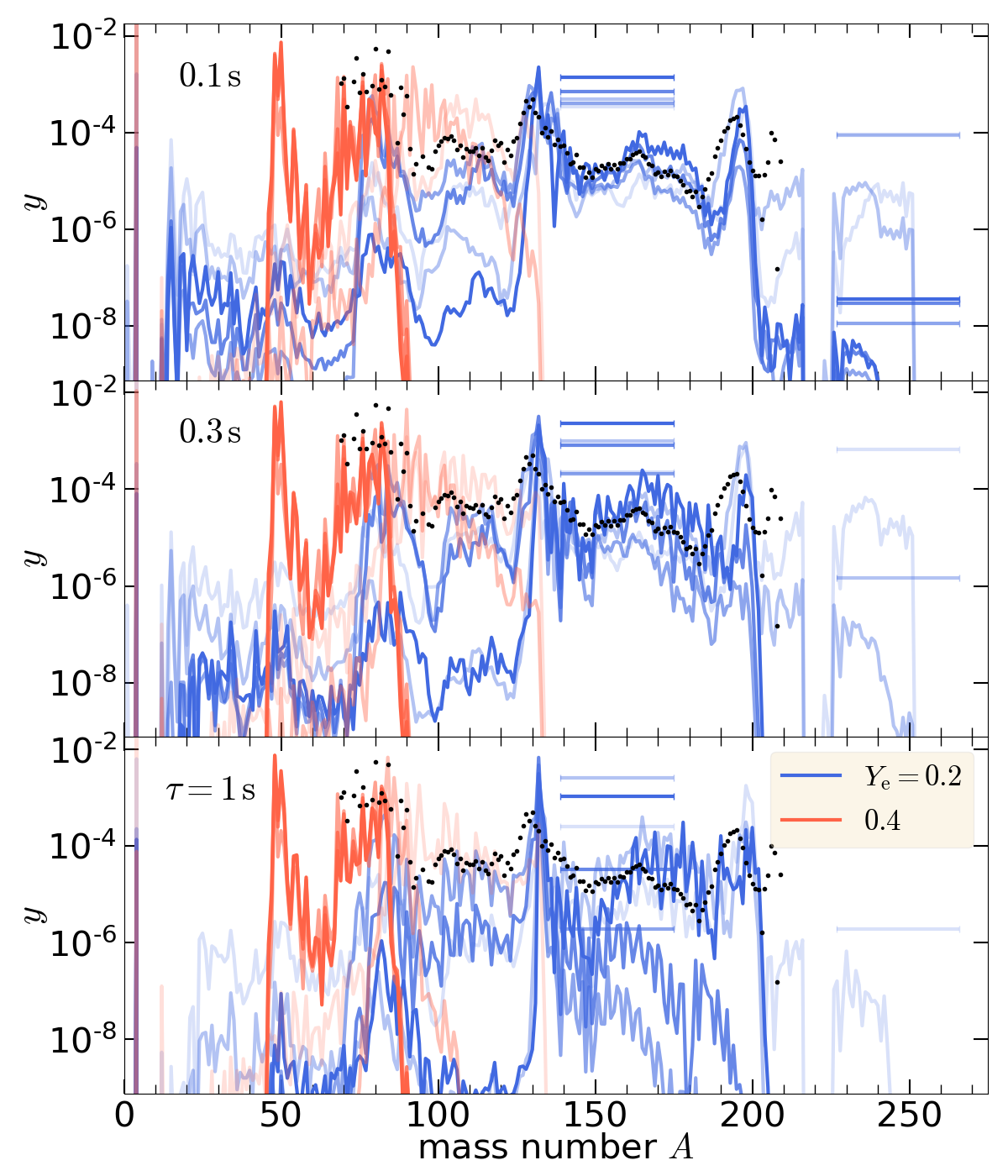}
\caption{The ``final'' (taken at $t=10^9\rm\, s$) mass fraction $y$ of different species of mass number $A$ for the nucleosynthesis cases in Fig. \ref{fig:heating_rates}. Here, only two initial electron fractions $\Ye=0.2$ (blue) and $0.4$ (red) are shown. Different shaded lines with the same color correspond to five logarithmically spaced initial entropies between $8$ and $350\kB/\mp$. The black dots are solar r-process elemental abundances \citep{arnould07_solar_abundance}, scaled to $y=5\times10^{-4}$ at $A=130$. The horizontal bars show the total mass fractions of Lanthanides ($139\leq A\leq 175$) and Actinides ($227\leq A\leq 266$).
}\label{fig:yfinal}
\end{figure}

We model the nucleosynthesis \textit{inside} the accretion disk using $\mathtt{SkyNet}$ \citep{lippuner17_skynet}. The system is initially in nuclear statistical equilibrium (NSE) at temperature $T_0=6\rm\, GK$ and specific entropy $s_0\in (8, 350)\kB/\mp$ (in units of the Boltzmann constant per proton mass). The specific entropy determines the initial density (for a radiation pressure-dominated gas, we have $s_0\propto T^3/\rho$) and the range of $s_0$ considered here spans from (electron-)degenerate gas and non-degenerate gas \citep[see Fig. 14 of][]{fernandez19_long_GRMHD}. We follow the adiabatic evolution of the system as it undergoes expansion on a timescale of $\tau=0.1$ to $1\rm\, s$. The results are shown in Fig. \ref{fig:heating_rates} (time-dependent heating rate) and \ref{fig:yfinal} (composition at $t=10^9\rm\,s$).

We find that the heating rates in the time range $10\lesssim t\lesssim 10^4\rm\, s$ do not depend sensitively on the initial conditions, provided that the initial $\Ye\lesssim 0.4$. The (neutrino subtracted) heating rate can be approximated by eq. (\ref{eq:qh}). We also find that for a longer expansion time $\tau$ (and hence neutron captures occur on a longer timescale), less Lanthanides and Actinides are produced.

\label{lastpage}
\end{document}